\documentclass[a4paper,11pt]{article}
\pdfoutput=1
\usepackage{jheppub}
\usepackage{graphicx}
\usepackage{dcolumn}
\usepackage{bm}
\usepackage{adjustbox}
\usepackage{multirow}
\usepackage{slashed}
\usepackage{hhline}
\usepackage{mathrsfs}
\usepackage[utf8]{inputenc}
\usepackage{tabularx,booktabs,caption}
\usepackage{cancel}
\usepackage{amssymb}
\usepackage{textcomp}
\usepackage{amsmath}
\usepackage[normalem]{ulem}
\usepackage{mathrsfs}
\usepackage{subfigure}
\usepackage[colorlinks=false,urlcolor=cambridgeblue,linkcolor=blue,citecolor=cambridgeblue,linktocpage=true,pdfproducer=medialab,pdfa=true,anchorcolor=blue]{hyperref}

\usepackage{empheq}
\usepackage{bm}
\usepackage{framed}
\usepackage{pdfpages}
\usepackage{subfigure}

\usepackage{slashed}
\usepackage{graphicx,color}
\usepackage{epsfig}
\usepackage{subfigure}
\usepackage{epsfig}
\usepackage{dcolumn}
\usepackage{bm}
\usepackage{color}
\usepackage{ulem}
\usepackage{enumitem}
\def\beq{\begin{equation}}
\def\eeq{\end{equation}}

\newcommand{\lsim}{\buildrel < \over {_\sim}}
\newcommand{\gsim}{\buildrel > \over {_\sim}}

\newcommand{\be}{\begin{equation}}
\newcommand{\ee}{\end{equation}}
\newcommand{\bea}{\begin{eqnarray}}
\newcommand{\eea}{\end{eqnarray}}
\newcommand{\ba}{\begin{array}}
\newcommand{\ea}{\end{array}} 
\newcommand{\tew}{T_\mathrm{EW}}
\newcommand{\shat}{\hat{s}}

\newcommand{\mrm}[1]{{\color{black}  #1}}

\begin{document}


\title{The Electroweak Phase Transition: A Collider Target}

\preprint{ACFI-T19-14}

\author[a,b,c]{Michael J. Ramsey-Musolf}
\emailAdd{mjrm@sjtu.edu.cn,mjrm@physics.umass.edu}

\affiliation[a]{Tsung-Dao Lee Institute, and School of Physics and Astronomy, Shanghai Jiao Tong University, Shanghai 200240, China}
\affiliation[b]{Amherst Center for Fundamental Interactions, University of Massachusetts-Amherst, Department of Physics, Amherst, MA 01003, USA}
\affiliation[c]{Kellogg Radiation Laboratory, California Institute of Technology, Pasadena, CA 91125 USA}

\abstract{
Determining the thermal history of electroweak symmetry breaking (EWSB) is an important challenge for particle physics and cosmology. Lattice simulations indicate that EWSB in the Standard Model (SM) occurs through a crossover transition, while the presence of new physics beyond the SM could alter this thermal history. The occurrence of a first order EWSB transition would be particularly interesting, providing the needed pre-conditions for generation of the cosmic matter-antimatter asymmetry and sources for potentially observable gravitational radiation. I provide simple, \mrm{generic} arguments that if such an alternate thermal history exists, the  new particles involved cannot be too heavy with respect to the SM electroweak temperature, nor can they interact too feebly with the SM Higgs boson. \mrm{These arguments do not rely on the decoupling limit. I derive } corresponding quantitative expectations for masses and interaction strengths \mrm{which} imply that their effects could in principle be observed (or ruled out) by prospective next generation high energy colliders. The simple, \mrm{generic} arguments \mrm{provide a quantitative, parametric understanding of} results obtained in a wide range of explicit model studies; \mrm{ relate them explicitly to the electroweak temperature; and delineate broad contours of collider phenomenology pertaining to a non-standard history of EWSB. } }

\maketitle

\section{Introduction}
\label{sec:intro}

What is the thermal history of electroweak symmetry-breaking  (EWSB)? While this question has been the subject of theoretical investigation for more than four decades, the discovery of the Standard Model-like Higgs boson puts it squarely in the spotlight for particle physics and cosmology. Results of non-perturbative lattice computations indicate that within the minimal Standard Model (SM), EWSB occurs through a crossover transition at a temperature $T\sim 100$ GeV. Had the Higgs-like scalar been lighter than $\sim 70-80$ GeV, the transition would have been a first order phase transition\cite{Kajantie:1996qd,Kajantie:1996mn,Gurtler:1997hr,Laine:1998jb,Csikor:1998eu,Aoki:1999fi}. The situation is analogous to what occurs in quantum chromodynamics (QCD). For sufficiently small baryon chemical potential $\mu_B$, as pertains to a purely SM early universe, the transition to the confined phase of QCD at $T\sim 100$ MeV was also of a crossover character. This theoretical result is consistent with experimental studies of heavy ion collisions  at the Relativistic Heavy Ion Collider (RHIC). Going forward, the RHIC beam energy scan hopes to identify the onset of a first order transition (or critical point) at non-zero $\mu_B$ (for recent overviews, see Refs.~\cite{Tlusty:2018rif,Bzdak:2019pkr}).

In the case of EWSB, there exists strong motivation to consider the possibility of an alternate thermal history as compared to the crossover transition of the minimal SM, even with the mounting evidence that the observed 125 GeV scalar is, indeed, the expected SM Higgs boson. Experimentally, LHC results to date have yet to preclude the possibility that the SM-like Higgs exists within an extended scalar sector, whose interactions could modify the thermal history of EWSB. Theoretically, a plethora of well-motivated particle physics models place the SM-like Higgs within such a beyond the Standard Model (BSM) setting. The resulting patterns of early universe EWSB could be considerably richer than in the minimal SM, a possibility suggested by Weinberg\cite{Weinberg:1974hy} and others\cite{Langacker:1980kd,Hammerschmitt:1994fn,Cline:1999wi,Profumo:2007wc,Patel:2012pi,Patel:2013zla,Blinov:2015sna,Meade:2018saz,Baldes:2018nel,Glioti:2018roy}. Perhaps, most compellingly, open problems in cosmology could find their solutions through such extended scalar sectors. In particular, the presence of a first order electroweak phase transition (EWPT) could provide the necessary \mrm{out-of-equilibrium} conditions for generation of the cosmic baryon asymmetry through electroweak baryogenesis (for a review and references, see Ref.~\cite{Morrissey:2012db}). If a first order EWPT were sufficiently strong, the resulting distortions of spacetime in the early universe would have produced relic gravitational waves (GWs) that one might observe in the LISA mission or future GW detectors (for a review, see Ref.~\cite{Weir:2017wfa}). In addition, additional neutral scalars could comprise part of the observed dark matter relic density. 

In this context, it is interesting to ask: What would it take experimentally to probe exhaustively the possibilities for the thermal history of EWSB and/or confirm that -- for all intents and purposes -- EWSB did in fact occur through a smooth crossover? In what follows, I present simple, \mrm{general} arguments for the vital role to be played by the LHC and future high energy colliders under active consideration. Importantly, the known properties of the SM electroweak interaction, together with the masses of the SM-like Higgs boson and the top quark, set the temperature scale for EWSB, or $\tew$. The arguments below imply that if there exist new bosons whose interactions substantially modify the thermal history of EWSB, their masses must be at most a few times $\tew$, making their possible existence a target for present and prospective colliders. For strongly coupled scenarios -- manifested as higher dimension operators in the Higgs potential -- the associated mass scale is similarly bounded. Specific, model-dependent studies performed to date are consistent with these general expectations. I also discuss the implications for precision measurements of Higgs boson properties. In this case, I obtain approximate lower bounds on the magnitude of deviations from SM Higgs properties that would follow from a significant departure from the SM EWSB thermal history. Although the arguments in the latter case are less airtight than those germane to the mass scale, they nevertheless provide a clear benchmark for precision Higgs boson studies that may be achievable with the LHC and prospective future colliders. In both cases (mass and precision), a wide range of EWPT studies involving specific model realizations are consistent with the general arguments provided below. \mrm{These realizations include Higgs portal models with gauge singlets -- either real\cite{Espinosa:1993bs,Benson:1993qx,Choi:1993cv,Vergara:1996ub,Ham:2004cf,Ahriche:2007jp,Profumo:2007wc,Noble:2007kk,Espinosa:2007qk,Espinosa:2008kw,Barger:2007im,Ashoorioon:2009nf,Das:2009ue,Espinosa:2011ax,Cline:2012hg,Chung:2012vg,Barger:2011vm,Huang:2012wn,Damgaard:2013kva,Fairbairn:2013uta,No:2013wsa,Profumo:2014opa,Craig:2014lda,Curtin:2014jma,Chen:2014ask,Katz:2014bha,Kozaczuk:2015owa,Kanemura:2015fra,Damgaard:2015con,Huang:2015tdv,Kanemura:2016lkz,Kotwal:2016tex,Brauner:2016fla,Huang:2017jws,Chen:2017qcz,Beniwal:2017eik,Cline:2017qpe,Kurup:2017dzf,Alves:2018jsw,Li:2019tfd,Gould:2019qek,Kozaczuk:2019pet,Carena:2019une,Heinemann:2019trx}  or complex\cite{Branco:1998yk,Barger:2008jx,Jiang:2015cwa,Chiang:2017nmu,Chiang:2019oms} -- or with electroweak multiplets\cite{FileviezPerez:2008bj,Chowdhury:2011ga,Patel:2012pi,Blinov:2015sna,Niemi:2018asa,Chao:2018xwz,Bell:2020gug,Chiang:2020rcv,Niemi:2020hto}; the two Higgs doublet model (2HDM)\cite{Turok:1991uc,Davies:1994id,Hammerschmitt:1994fn,Cline:1996mga,Fromme:2006cm,Cline:2011mm,Dorsch:2013wja,Dorsch:2014qja,Harman:2015gif,Basler:2016obg,Dorsch:2017nza,Bernon:2017jgv,Andersen:2017ika,Kainulainen:2019kyp}; the minimal supersymmetric SM (MSSM)\cite{Carena:1996wj,Delepine:1996vn,Cline:1996cr,Laine:1998qk,Carena:2008vj,Cohen:2012zza,Laine:2012jy,Curtin:2012aa,Carena:2012np,Katz:2015uja} and its extensions with gauge singlet superfields\cite{Pietroni:1992in,Davies:1995un,Huber:2000mg,Ham:2004nv,Ham:2004pd,Menon:2004wv,Funakubo:2005pu,Huber:2006ma,Chung:2010cd,Kozaczuk:2014kva,Huang:2014ifa}; and the SM effective field theory with dimension six operators in the Higgs potential\cite{Grojean:2004xa,Grinstein:2008qi,Huang:2015izx,Cao:2017oez}.} 

\mrm{ It may occur to the reader that when a new particle interacting with the Higgs boson becomes sufficiently massive, it will decouple from the thermal bath, thereby yielding the SM EWSB thermodynamics. What may be less obvious -- and what I discuss in generality below -- is that one need not utilize the onset of this decoupling limit
in order to obtain quantitative mass and precision bounds. Indeed, most of the following discussion relies on explicit retention of new degrees of freedom in the theory at non-zero temperature. Moreover, the rich array of EWSB patterns that can emerge in this regime preclude any statements on the level of a theorem. Rather, one must examine the possibilities in as simple and generic a manner so as to avoid any impression that the results from model studies collectively leave open significant loopholes. 

Thus, in what follows, I lay out such a simple and generic framework; show it how leads to generally-applicable quantitative benchmarks; connect the latter explicitly to $\tew$; and use it to delineate the broad contours of collider phenomenology pertaining to the thermal history of EWSB. Wherever possible, I also make explicit reference to results existing model studies, which can be quantitatively understood in terms of this simple framework, and which are generally not expressed directly in terms of $\tew$. Moreover, in a wide subset of these model studies, the choice of viable parameters often conflate two criteria: (a) the generation of a first order EWPT and (b) the requirement that it be be a sufficiently \lq\lq strong" transition so as to facilitate EWBG and/or observable gravitational wave production. In what follows, I draw a clear distinction between these two criteria and the resulting implications for collider phenomenology.
In this respect, it is not {\em a priori} clear that the LHC and prospective future collider sensitivities are well-matched to achieving a comprehensive probe of the landscape of EWSB thermal histories in SM extensions. The conclusion that they are -- as a direct consequence of the scale $\tew$ -- is a primary finding of this work.}

\mrm{Before proceeding, it may help the reader to better appreciate the spirit and novelty of this work by making analogy with studies in another context, namely, so-called \lq\lq natural supersymmetry(SUSY)" \cite{Papucci:2011wy}. Natural  SUSY models impose restrictions on parameters most relevant to solving the hierarchy problem without excessive fine tuning, while allowing wider flexibility for other parameters in the soft SUSY-breaking Lagrangian (for early realizations see, {\em e.g.} Refs.~\cite{Dimopoulos:1995mi,Cohen:1996vb}). The resulting predictions for experimental signatures -- based on a specific scale (the weak scale) and a theoretical requirement (the absence of fine-tuning) -- are then rather generic (for reviews and applications to LHC results, see, {\em e.g.} Refs.~\cite{Feng:2013pwa,Craig:2013cxa,Evans:2013jna}). Special exceptions may, of course, still occur, and such examples are interesting in their own right. In the present study, I ask an analogous question: What are the generic, quantitative expectations for new scalar masses and for deviations of Higgs boson properties in models that qualitatively alter the SM thermal history of EWSB? The framework discussed below addresses these questions by making reference to a specific scale ($\tew$) and and theoretical requirement (altered thermal EWSB history) while remaining as agnostic as possible about the complete theory that contains its key ingredients.  In a few special model realizations, it may be possible to find regions of parameter space where one can evade the general arguments that follow from this framework.  These interesting cases are exceptions that prove the rule, and in some instances, may be subject to other experimental probes such as dark matter direct detection (see, {\em e.g.}, Refs.~\cite{Curtin:2014jma,Chiang:2017nmu}). As the focus of this work is on the generally applicable considerations that apply to a broad swath of specific models, I will not treat these exceptional cases here.}

The discussion of these \mrm{general arguments and their implications} is organized as follows. In Section \ref{sec:ewtemp}, I review the basic features of SM EWSB at non-zero temperature, using a simple perturbative framework in the high temperature effective theory. I then consider mechanisms by which new bosonic degrees of freedom having renormalizable interactions may modify this thermal history. The quantitative implications for BSM mass scale and Higgs boson properties are considered in Sections \ref{sec:mass} and \ref{sec:higgsprop}, respectively. Section \ref{sec:collider} gives a brief overview of the relevant mass reach and precision at the LHC and prospective future colliders. \mrm{Representative benchmarks appear in Fig.~\ref{fig:hgg} and Tables~\ref{tab:epem}-\ref{tab:ppsinglet}.}In Section \ref{sec:other}, I discuss scenarios with non-renormalizable operators in the Higgs potential and scenarios involving new light degrees of freedom. In \ref{sec:tewrev}, 
I return to the consideration of $\tew$, discussing possible modifications of this scale due to BSM scalar interactions, \mrm{including non-renormalizable operators}. 
Some perspective and outlook appear in Section \ref{sec:out}.  

\section{The Electroweak Temperature}
\label{sec:ewtemp}

The starting point for this discussion is the SM Higgs potential
\be
V_0(H) = -\mu^2 H^\dag H + \lambda (H^\dag H)^2\ \ \ .
\label{eq:v0H}
\ee

At non-zero $T$, the dynamics of EWSB are driven by the finite-temperature effective potential, expressed as a function of the $h= \sqrt{2}\mathrm{Re} (H^0)$, where $\langle h \rangle \equiv v=246$ is the $T=0$ Higgs vacuum expectation value (vev):
\be
V(h,T)_\mathrm{eff} ^\mathrm{SM}= V_0(h) + V_\mathrm{CW} (h) + V(h, T)_\mathrm{SM}\ \ \ ,
\label{eq:vh}
\ee
where $V_\mathrm{CW} (h)$ is the zero temperature Coleman-Weinberg potential and where $V(h, T)_\mathrm{SM}$ is generated by thermal loops (for a pedagogical introduction, see, {\it e.g.}, Ref.~\cite{Quiros:1999jp}). For present purposes it is convenient to consider the latter in the high-temperature limit at leading order (LO) and to neglect (momentarily) the $V_\mathrm{CW}$ contribution\footnote{We will return to its implications below.}, leading to
\be
\label{eq:veffsmhighT}
V(h, T)^\mathrm{SM}_\mathrm{eff} = D(T^2-T_0^2)\, h^2 +\frac{\lambda}{4}\, h^4\ \ \ ,
\ee
where we have truncated at second order in the EW gauge couplings at one-loop in order to avoid complications associated with gauge-dependence that are not germane to the present discussion (see Refs.~\cite{Patel:2011th,Laine:1994zq,Ekstedt:2018ftj,Ekstedt:2020abj} and references therein). We have also dropped terms that depend logarithmically on $T$ or that fall off as inverse powers of the temperature. The quantity $D$ is calculable in perturbation theory, while $DT_0^2 = \mu^2/2$. The $T=0$ EWSB minimization conditions  allow one to express $\mu^2$ in terms of $v$ and the Higgs quartic self coupling, yielding 
\be
\label{eq:T0sq}
T_0^2 = \left(8 \lambda +  \ \mathrm{loops} \right)\left(4\lambda+ \frac{3}{2} g^2 + \frac{1}{2} g^{\prime\ 2} + 2 y_t^2 + \cdots\right)^{-1}\ v^2\ \ \ .
\ee
Here, we have retained only the dominant, leading contributions to the temperature $T_0$. From the experimental values for the quantities on the RHS of Eq.~(\ref{eq:T0sq}) we obtain the EW temperature
\be
\label{eq:TEWvalue}
T_\mathrm{EW} \equiv T_0 \approx 140\ \mathrm{GeV}\ \ \ .
\ee
At this stage, two points merit emphasis. First, the foregoing result has been obtained in a simple, perturbative framework. In this context, the nature of the EWSB transition implied by Eq.~(\ref{eq:veffsmhighT}) is a second order phase transition. The critical temperature for the latter is $T_0$, below which the quadratic term becomes negative and the minimum of energy lies away from the symmetric phase ($h=0$). Lattice studies indicate that transition is not a true second order phase transition, as there exists no evidence for diverging correlation lengths when the Higgs vev becomes non-zero. Nevertheless, for lighter values of $m_H$ consistent with a first order transition,  the lattice values for the transition temperature are consistent with those obtained using Eq.~(\ref{eq:TEWvalue}) (see, {\it e.g.}, Ref.~\cite{Kajantie:1996mn}). Thus, it suffices for our purposes to take $\tew\approx 140$ GeV (we return to a more detailed discussion in Section \ref{sec:tewrev}).

Second, the measured values of the weak scale (equivalently, the muon lifetime), Higgs boson and top quark masss -- yielding $\lambda$ and $y_t$, respectively, and EW gauge couplings are all decisive inputs in setting the value of $\tew$. It would be a misconception to identify $\tew$ with $v$, even though they are numerically commensurate. Had the experimental Higgs and/or top quark masses in particularly taken on different values, one might have obtained an electroweak temperature appreciably different from $v$. In short, one should consider $\tew$ and $v$ to be distinct scales. 

We now ask: how might the presence of new particles and interactions or strong dynamics modify the foregoing picture? And what are the implications for the corresponding mass scale $M$ ? 

The impact of new particles and interactions occurs via one or more for the following avenues:
\begin{itemize}
\item Thermal loops
\item Modification of the tree-level $T=0$ vacuum structure
\item Loop induced modification of the $T=0$ vacuum 
\end{itemize}
We consider the first two of these possibilities in turn, and return to the third in Section \ref{sec:tewrev}. For purposes of concrete \mrm{yet general} illustration, we introduce an additional scalar field $\phi$ that can transform either as a singlet under S(3$)_C\times$SU(2$)_L\times$U(1$)_Y$ or as a non-trivial representation of one or more of these factors. For non-singlet representations, consider the following gauge invariant, renormalizable $T=0$ potential 
\be
V_0(H,\phi) = V_0(H) + V_0(\phi) + V_0(H,\phi)
\ee
with
\begin{subequations}
\begin{align}
\label{eq:v0phi}
V_0(\phi) & =  \frac{b_2}{2} \phi^\dag\phi +\frac{b_4}{4!} (\phi^\dag\phi)^2\\
\label{eq:v0hphi}
V_0(H,\phi) &=   \frac{a_2}{2} \phi^\dag\phi H^\dag H \ \ \ .
\end{align}
\end{subequations}
In general $\phi$ may be complex. When $\phi$ is a SM gauge singlet, it may be either real or complex. In this case, both $V_0(\phi) $ and $V_0(H,\phi)$ may contain terms that are odd under $\phi\to -\phi$. For $\phi$ transforming as a singlet or triplet under SU(2$)_L$, one may also include cubic interaction terms of the form $H^\dag\phi H$. We return to these possibilities in Section \ref{sec:higgsprop}. We also note that for higher dimensional representations of the SM gauge groups, the interactions in Eqs.~(\ref{eq:v0phi}, \ref{eq:v0hphi}) give a simplified, symbolic expression for the full set of allowed operators. For a complete discussion in the case of higher dimensional representations of SU(2$)_L\times$U(1$)_Y$, see Ref.~\cite{Chao:2018xwz}. For our present purposes of \mrm{assessing in a general way} the impact on phase transition dynamics and the collider reach, the simplified model forms in Eqs.~(\ref{eq:v0phi}, \ref{eq:v0hphi}) suffice. 


\begin{figure}
\begin{center}
  \includegraphics[width=0.8\textwidth]{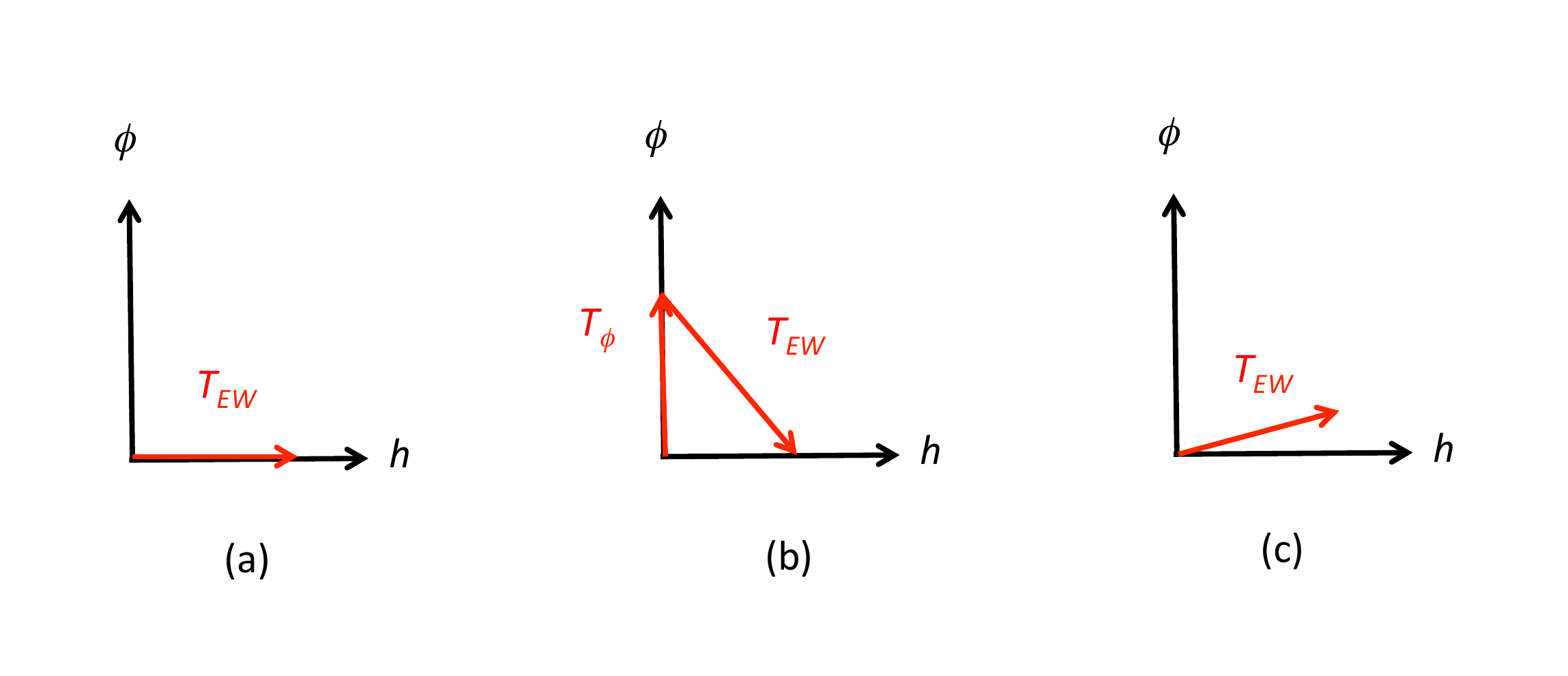}
\end{center}
\caption{Representative thermal histories of EWSB in the presence of the Higgs field $h$ and additional neutral scalar $\phi$. (a) A one-step transition to the pure Higgs phase at temperature $\tew$; (b) a two-step transition, with a first step to the $\phi$-vacuum at $T_\phi > \tew$ followed by a second step to the pure Higgs phase; (c) a one-step transition at temperature $\tew$ to the a mixed phase in which both $h$ and $\phi$ have non-zero vacuum expectation values.\label{fig:scenarios}}
\end{figure}


With this simple framework in mind, it is useful to consider various possible thermal histories of EWSB as illustrated in Fig.~\ref{fig:scenarios}. Panels (a) and (b) are most relevant to the form of the $H$-$\phi$ interaction given in Eq.~(\ref{eq:v0hphi}) whereas the trajectory of panel (c) is of interest when additional terms containing odd powers of $\phi$ are included (see below). Scenario (a) is, perhaps, the most familiar one, corresponding to a single-step transition to the present pure Higgs vacuum from the symmetric phase at $T=\tew$. Scenario (b) illustrates a two-step transition history, wherein the universe first lands in a phase with a non-zero $\phi$ vacuum expectation value at $T=T_\phi>\tew$ followed by a transition to the  pure Higgs vacuum at $\tew$. For scenario (c) the final vacuum is one in which both the Higgs and neutral component of $\phi$ obtain vevs. The transition shown occurs in a singlet step, though it is also possible for a multi-step transition to lead to the same vacuum.

\section{Mass Reach}
\label{sec:mass}

\vskip 0.2in

\noindent {\em Thermal loops}. The most significant, finite $T$ loop effects of the new interactions given above are additional contributions to the Higgs boson thermal mass (the $D$ coefficient) and the possible introduction of a barrier between the broken and unbroken vacua, opening up the possibility of a first order EWPT. We focus here on this possibility in the context of the scenario in Fig.~\ref{fig:scenarios}(a). The relevant contribution to $V(h, T)$, obtained after performing the \lq\lq daisy resummation" (see \cite{Quiros:1999jp} for a discussion and references), is
\be
 \label{eq:DeltaV}
\Delta V(h,T) \supset -\frac{T}{12\pi} M_\phi(h,T)^3
\ee
with
\be
\label{eq:mphiT}
M_\phi(h,T)^3 = \left[\left(\frac{a_2}{6}+\frac{b_4}{4}+\Delta_g\right) T^2 + b_2 + \frac{a_2}{2} h^2\right]^{3/2}\ \ \ .
\ee
The first term inside the RHS gives the thermal mass-squared of the $\phi$ arising from its interactions with $H$ and itself and from possible gauge loops ($\Delta_g$).  The second and third terms will give the square of the $T=0$ mass of the $\phi$ when $h\to v$. Consider now the possibility that the first two terms cancel for $T\sim \tew$:
\be
b_2 + \left(\frac{a_2}{6}+\frac{b_4}{4}+\Delta_g\right) T^2_\mathrm{EW} \approx 0
\label{eq:loopcancel}
\ee
One may ensure tree-level stability of the full potential by choosing $a_2>0$, $b_4>0$,  and $b_2<0$. In this case, one has
\be
\Delta V(h,T_\mathrm{EW}) \supset -\frac{T_\mathrm{EW}}{12\pi} \frac{a_2^{3/2}}{2\sqrt{2}} h^3 \ \ \ .
\label{eq:loopcubic}
\ee
The cubic dependence on $h$ constitutes a barrier between the symmetric and broken phases along the Higgs direction, implying that the transition to the broken phase will be a first order EWPT. We now consider the implications for the $\phi$ mass at $T=0$. It is useful to first consider the $b_4\to 0$ and $\Delta_g\to 0$ limit:
\be
M_\phi^2(T=0) = \frac{a_2}{2}\left(v^2-T_\mathrm{EW}^2/3\right)\ \ \ .
\label{eq:mphiloop}
\ee
We observe that the scale of $M_\phi(T=0)$ depends crucially on the difference between $v$ and $\tew/\sqrt{3}$ and on the magnitude of the Higgs portal coupling. Varying $a_2$ between unity and five leads to the following range for the $\phi$ mass:
\be
170\ \mathrm{GeV} \lsim M_\phi(T=0) \lsim 375\ \mathrm{GeV}\ \ \ .
\label{eq:mphibound}
\ee
Including a non-zero value for $b_4$ reduces the value of $M_\phi(T=0)$, so the upper end of the range is, in some sense, maximal for the foregoing inputs. That being said, there exist several reasons for relaxing this upper bound somewhat:
\begin{itemize}
\item The criteria for defining perturbative values of the coupling is somewhat ambiguous. Use of the na\" ive perturbativity bounds $a_2< 2\pi$ and $b_4< 2\pi/3$ leads to $\sim 10\%$ larger maximum  $M_\phi(T=0)$. According to the work of Ref.~\cite{Riesselmann:1996is}, which considered the perturbativity constraints on the Higgs quartic self-coupling, these na\" ive $T=0$ perturbativity bounds may be too restrictive, implying that larger values of the couplings may be consistent with perturbativity. On the other hand, the importance of higher-dimensional non-renormalizable operators in the finite-$T$ effective potential will also grow with the values of the couplings.  Any analysis performed in this region of parameter space would need to include the effects of these operators before making definitive statements. 
\item As discussed further below, including the contribution of the cross-quartic interaction will introduce a linear dependence on $a_2$ in the denominator of Eq.~(\ref{eq:T0sq}) at one-loop order, thereby lowering $\tew$ and increasing $M_\phi(T=0)$. 
\item A first order transition may arise even if the cancellation in Eq.~(\ref{eq:loopcancel}) is not exact. 
\end{itemize}
With these considerations in mind, I will take 700 GeV to be a conservative upper bound for the new scalar mass.

\mrm{In general,  this $T>0$ loop effect can be present in all models containing the portal interaction $\phi^\dag\phi H^\dag H$. Its impact has been most thoroughly studied in the MSSM, where $\phi$ is the scalar partner of the right-handed (RH) top quark. In this case, the requirements of supersymmetry imply that  $a_2\sim y_t^2$ and $b_4\sim 4\pi\alpha_s$, up to corrections associated with mixing between the left- and right-handed stops. In order to obtain the observed value of the Higgs boson mass in the MSSM, the left-handed stop must be several orders of magnitude larger than the 700 GeV benchmark obtained here\cite{Curtin:2012aa,Carena:2012np}, leaving open only the RH stop as a candidate for catalyzing this loop effect. The corresponding upper bound on the RH stop mass-- roughly 100-150 GeV\cite{Carena:2008vj,Laine:2012jy} -- is well below the conservative benchmark obtained here, owing to the details of the stop mass matrix, the value of the Higgs boson mass,  and requirements of SUSY.  In this context it is worth noting that the number of RH stop degrees of freedom associated with its charge under QCD does not change the argument for the mass scale in Eqs.~(\ref{eq:loopcancel}-\ref{eq:mphiloop}). Rather, one simply multiplies the RHS of Eq.~(\ref{eq:loopcubic}) by the number of colors, $N_C$. A similar comment applies to scalars $\phi$ that exist in other representations of SU(3$)_C$ or SU(2$)_L$. The dimensionality of the representation is not directly decisive for the mass bound, but does impact the over height of the barrier and the corresponding \lq\lq strength" of the EWPT (see below). 

For the specific case of the MSSM with the light RH stop, early LHC measurements of Higgs boson properties, coupled with constraints from electroweak precision tests, effectively exclude this scenario\cite{Curtin:2012aa}, even if tensions with Higgs boson data are eased by introduction of a hard SUSY-breaking term in the scalar potential\cite{Katz:2015uja}. For other scenarios, such as the generic (non-supersymmetric) Higgs portal models with gauge singlets\cite{Espinosa:1993bs,Benson:1993qx,Choi:1993cv,Vergara:1996ub,Branco:1998yk,Ham:2004cf,Ahriche:2007jp,Profumo:2007wc,Noble:2007kk,Espinosa:2007qk,Espinosa:2008kw,Barger:2007im,Barger:2008jx,Ashoorioon:2009nf,Das:2009ue,Espinosa:2011ax,Cline:2012hg,Chung:2012vg,Barger:2011vm,Huang:2012wn,Damgaard:2013kva,Fairbairn:2013uta,No:2013wsa,Profumo:2014opa,Craig:2014lda,Curtin:2014jma,Chen:2014ask,Katz:2014bha,Kozaczuk:2015owa,Kanemura:2015fra,Damgaard:2015con,Huang:2015tdv,Jiang:2015cwa,Kanemura:2016lkz,Kotwal:2016tex,Brauner:2016fla,Huang:2017jws,Chiang:2017nmu,Chen:2017qcz,Beniwal:2017eik,Cline:2017qpe,Kurup:2017dzf,Alves:2018jsw,Li:2019tfd,Gould:2019qek,Kozaczuk:2019pet,Carena:2019une,Heinemann:2019trx} or color-neutral electroweak multiplets\cite{FileviezPerez:2008bj,Chowdhury:2011ga,Patel:2012pi,Blinov:2015sna,Niemi:2018asa,Chao:2018xwz,Bell:2020gug,Chiang:2020rcv,Niemi:2020hto}, the constraints from Higgs boson properties and  electroweak precision tests are not nearly as severe, allowing for a significantly larger value of $M_\phi(T=0)$ consistent with present phenomenological constraints. For example, the recent non-perturbative analysis for $\phi$ transforming as a real, electroweak triplet in Ref.~\cite{Niemi:2018asa}-- based on the dimensionally-reduced finite $T$ effective field theory and earlier lattice computations -- indicates that a single step, thermal loop-induced first order transition to the SM electroweak vacuum (Fig.~\ref{fig:scenarios}a) can occur for $M_\phi(T=0)$ as large as $\sim 500$ GeV for a portal coupling $a_2$ near the upper end of the nominal perturbative range I have used here. 
At the same time,  changes in the tree-level vacuum structure can open significant new avenues for a first order EWPT in these general Higgs portal models, as well as in singlet extensions of the MSSM or scenarios with non-renormalizable operators in the scalar potential.
}

\vskip 0.2in

\noindent {\em Tree-level vacuum structure}. A second possibility for changing the thermal history of EWSB involves a multi-step transition into the Higgs phase, as in Fig.~\ref{fig:scenarios}(b).  During the first step the universe goes to a vacuum in which $\langle\phi^0\rangle\not=0$, followed by a second step to the Higgs phase. We will refer to the corresponding critical temperatures as $T_\phi$ and $T_h$, respectively.  In this case, the operator $\phi^\dag\phi H^\dag H$ creates a barrier between the $\phi$ and Higgs vacua if the coefficient $a_2/2>0$. 

The two step history requires $T_\phi > T_h$. We will again take $T_h\approx \tew$. The value of $T_\phi$ is determined from the quadratic term in the part of the potential independent of $H$:
\be
V(\varphi, T) = \frac{1}{2} \left[-|b_2| + \frac{T^2}{2} \left(\frac{a_2}{3}+\frac{1}{2} b_4 + 2\Delta_g\right)\right] \varphi^2 + \frac{b_4}{4!} \varphi^4\ \ \ ,
\ee
where, as before, we have retained only the leading $T$-dependent terms. Setting the coefficient of the quadratic term to zero implies that
\be
|b_2| = \frac{T^2_\phi}{6} \left(a_2+\frac{3}{2} b_4+ 6\Delta_g\right) > \frac{T^2_\mathrm{EW}}{6} \left(a_2+\frac{3}{2} b_4+ 6\Delta_g\right)\ \ \ ,
\ee
where the inequality follows from the requirement that $T_\phi > T_h\approx \tew$. 
Now compute the mass $T=0$ $\phi$ mass:
\be
M_\phi(0) = \left[\frac{a_2}{2}v^2 - |b_2|\right]^{1/2} < \left[\frac{a_2}{2}v^2-\frac{T^2_\mathrm{EW}}{6} \left(a_2+\frac{3}{2} b_4+ 6\Delta_g\right)\right]^{1/2} \ \ \ .
\ee
For illustration, consider $b_4=0.3$, neglect $\Delta_g$, and vary $a_2$ between one and five. Doing so leads to the range $160\ \mathrm{GeV} \lsim M_\phi(T=0) \lsim 360\ \mathrm{GeV}$, commensurate with the range associated with the thermal loop effect. Increasing the value of $b_4$ will lead to lower values of $M_\phi(T=0) $, while taking the unphysical  $b_4\to 0$ limit will lead only to a slight increase. As in the previous argument, I will take 700 GeV as a conservative upper bound, taking into account the ambiguities in defining the requirements of perturbativity the restriction to renormalizable interactions.

It is worth noting that the if $\phi$ carries electroweak charge, then EWSB will occur twice in this scenario\cite{Hammerschmitt:1994fn,FileviezPerez:2008bj,Chowdhury:2011ga,Patel:2012pi,Blinov:2015sna,Niemi:2018asa,Chao:2018xwz,Bell:2020gug,Chiang:2020rcv,Niemi:2020hto}. The transition to the $\phi$ phase may itself be either a true phase transition (first or second order) or crossover. \mrm{For a recent analysis in the real triplet extension of the SM using effective theory and lattice simulations, see Ref.~\cite{Niemi:2020hto}.} On the other hand, for $\phi$ being either a SM gauge singlet or electroweak multiplet, the EWSB transition to the Higgs phase will be first order due to the presence of the tree-level barrier. \mrm{The latter possibility has been studied explicitly in detail in SM extensions with gauge singlets \cite{Profumo:2007wc,Espinosa:2011ax,Curtin:2014jma,Chiang:2017nmu} and real triplets\cite{Patel:2012pi,Blinov:2015sna,Niemi:2020hto}. The resulting mass ranges are generally consistent with the general arguments given here }. \mrm{Importantly, the possibility of multi-step transitions open up} one or more new avenues for electroweak baryogenesis and gravitational wave generation \mrm{ (see, {\em e.g.}, Refs~\cite{Patel:2012pi,Inoue:2015pza,Jiang:2015cwa,Croon:2018new,Bell:2019mbn}.)}. 

\section{Higgs Boson Properties}
\label{sec:higgsprop}

In addition to performing direct searches for $\phi$ one may discern its interactions with the Higgs boson indirectly through precision measurements of Higgs boson properties. Here, we consider two categories of effects: (a) loop-induced modifications of Higgs properties arising from the cross-quartic interaction in Eq.~(\ref{eq:v0hphi}) and (b) tree-level modifications arising from $\phi$-$H$ interactions that break the $Z_2$ symmetry of the potential in (\ref{eq:v0phi},\ref{eq:v0hphi}). 

\subsection{$Z_2$-symmetric interactions}

\vskip 0.25in

\noindent{\em Higgs di-photon decays}. From the discussion in Section \ref{sec:mass}, it is evident that $a_2$ plays a decisive role in both generating the presence of a barrier in the finite-$T$ potential and in determining the magnitude of $M_\phi(T=0)$.  In order to test the validity of the foregoing arguments and more firmly establish the presence of the barrier, would be of interest to have additional information on $a_2$ and $M_\phi$. In this context, a precision measurement of the rate for the Higgs-like scalar to decay to two photons provides an interesting sensitivity to these two parameters when $\phi$ carries electroweak charge, \mrm{as in the case of the 2HDM\cite{Turok:1991uc,Davies:1994id,Hammerschmitt:1994fn,Cline:1996mga,Fromme:2006cm,Cline:2011mm,Dorsch:2013wja,Dorsch:2014qja,Harman:2015gif,Basler:2016obg,Dorsch:2017nza,Bernon:2017jgv,Andersen:2017ika,Kainulainen:2019kyp} or general electroweak multiplets\cite{FileviezPerez:2008bj,Chowdhury:2011ga,Patel:2012pi,Blinov:2015sna,Niemi:2018asa,Chao:2018xwz,Bell:2020gug,Chiang:2020rcv,Niemi:2020hto}.} Assuming, for example, that one determines $M_\phi$ by other means,   such as direct production (see below), the $\Gamma(h\to\gamma\gamma)$ provides a probe of $a_2$.  

For the general Higgs portal interaction of Eq.~(\ref{eq:v0hphi}), the relative change in $\Gamma(h\to\gamma\gamma)$ is given to a good approximation by
\be
\frac{\Delta\Gamma(h\to\gamma\gamma)}{\Gamma(h\to\gamma\gamma)^\mathrm{SM}} \approx 
\frac{|N_c q_t^2 F_{1/2}(\tau_t)+q_W^2 F_1(\tau_W) + \kappa_\phi (\sum_j q_{S_j}^2) F_0(\tau_S)|^2}{|N_c q_t^2 F_{1/2}(\tau_t)+q_W^2 F_1(\tau_W)|^2}-1
\ee
where $N_c=3$, $q_t$ and $q_W$ are top quark and W-boson charges (in units of $e$); $q_{S_j}$ is the charge of the $j$-th charged element of the $\phi$ mulitplet; $F_{0,1/2,1}(\tau_S)$ are well-known loop functions\cite{Gunion:1989we} of the parameter
\be
\tau_k = 4 m_j^2/m_h^2 \ \ \ ;
\ee 
\be
\kappa_\phi = 2 a_2 \times \left(\frac{v}{2 M_\phi}\right)^2\ \ \ ;
\ee
and where we have included only the top quark and W-boson contributions to $\Gamma(h\to\gamma\gamma)^\mathrm{SM}$. The invariant amplitude for the SM and $\phi$ contributions carry opposite (same) signs for positive (negative) $a_2$. 

\begin{figure}[t]
\begin{center}
\includegraphics[width=9cm,height=7.5cm,angle=0]{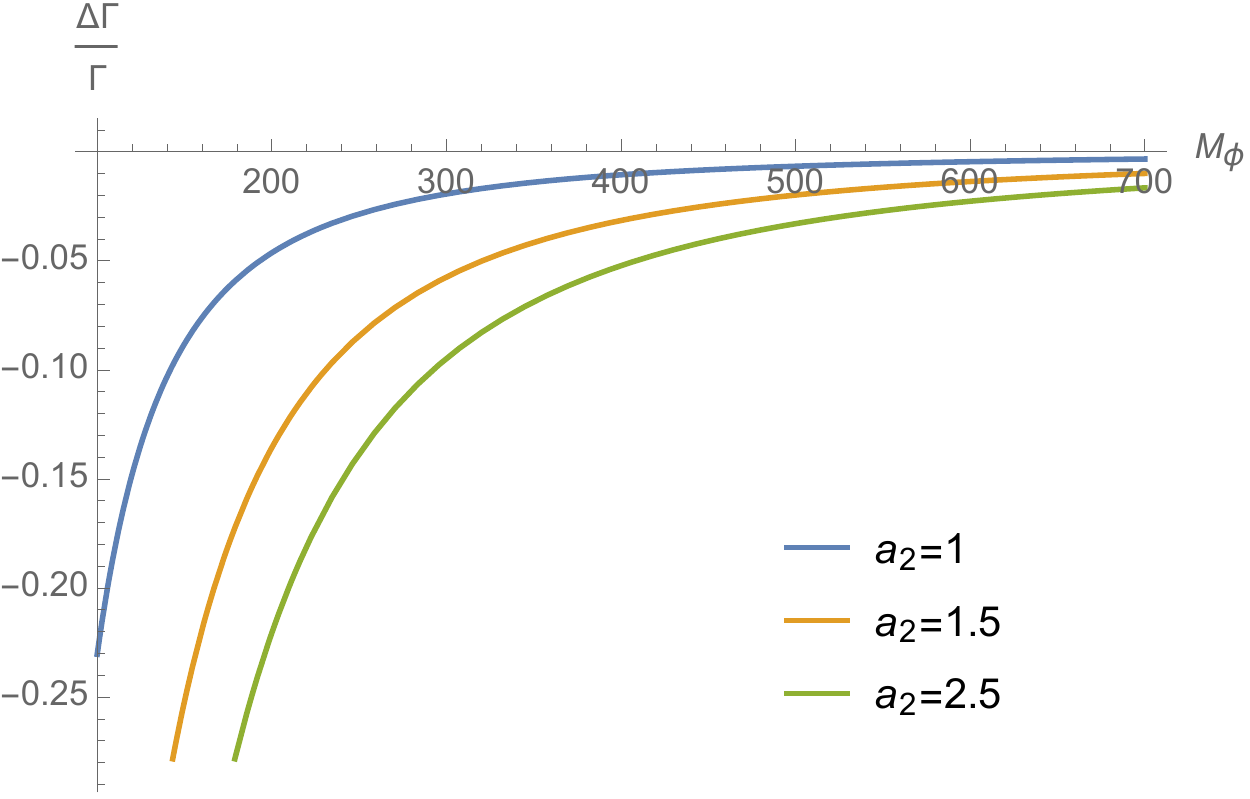}
\end{center}
\caption{Relative change in the Higgs di-photon decay rate induced by a scalar electroweak multiplet and the interaction in Eq.~(\ref{eq:v0hphi}) .  A first order EWPT to the Higgs vacuum in the scenarios of Fig.~\ref{fig:scenarios}(a,b) requires $a_2>0$, leading to a reduction in the di-photon decay rate. The different curves correspond to different representative choices for $a_2$.  }  \label{fig:hgg}\end{figure}

How large might one expect the magnitude of $\Delta\Gamma(h\to\gamma\gamma)/\Gamma(h\to\gamma\gamma)^\mathrm{SM}$ to be? In Fig. \ref{fig:hgg}, we give $\Delta\Gamma(h\to\gamma\gamma)/\Gamma(h\to\gamma\gamma)^\mathrm{SM}$ as a function of $M_\phi$ for representative values of $a_2$. Two important features emerge. First, the presence of a barrier driven by the cross-quartic Higgs portal interaction ($a_2>0$)  will reduce the di-photon decay rate relative to its SM value. Second, we observe that $\Delta\Gamma/\Gamma\sim\mathcal{O}(0.01)$ for $a_2\sim\mathrm{O}(1)$ and for $M_\phi$ near the upper end of the EWPT-viable mass range for $M_\phi$. For lighter masses (consistent with the LEP bounds), the effect can be on the order of ten percent.

\subsection{$Z_2$-breaking interactions}
\label{sub:z2break}

We now consider the possible inclusion of renormalizable operators that break the $Z_2$-symmetry of Eqs.~(\ref{eq:v0phi},\ref{eq:v0hphi}):
\be
\label{eq:z2break}
\Delta V_0(H, \phi) = \frac{b_3}{3!} \phi^3 + \frac{a_1}{2} H^\dag \phi H +\mathrm{h.c.}\ \ \ ,
\ee
where $\phi$ may be a \mrm{real SM gauge singlet\cite{Espinosa:1993bs,Benson:1993qx,Choi:1993cv,Vergara:1996ub,Ham:2004cf,Ahriche:2007jp,Profumo:2007wc,Noble:2007kk,Espinosa:2007qk,Espinosa:2008kw,Barger:2007im,Ashoorioon:2009nf,Das:2009ue,Espinosa:2011ax,Cline:2012hg,Chung:2012vg,Barger:2011vm,Huang:2012wn,Damgaard:2013kva,Fairbairn:2013uta,No:2013wsa,Profumo:2014opa,Craig:2014lda,Curtin:2014jma,Chen:2014ask,Katz:2014bha,Kozaczuk:2015owa,Kanemura:2015fra,Damgaard:2015con,Huang:2015tdv,Kanemura:2016lkz,Kotwal:2016tex,Brauner:2016fla,Huang:2017jws,Chen:2017qcz,Beniwal:2017eik,Cline:2017qpe,Kurup:2017dzf,Alves:2018jsw,Li:2019tfd,Gould:2019qek,Kozaczuk:2019pet,Carena:2019une,Heinemann:2019trx}; or complex gauge singlet -- supersymmetric \cite{Pietroni:1992in,Davies:1995un,Huber:2000mg,Ham:2004nv,Ham:2004pd,Menon:2004wv,Funakubo:2005pu,Huber:2006ma,Chung:2010cd,Kozaczuk:2014kva,Huang:2014ifa} or otherwise\cite{Branco:1998yk,Barger:2008jx,Jiang:2015cwa,Chiang:2017nmu,Chiang:2019oms}; or a real triplet that transforms as $(1,3,0)$ \cite{FileviezPerez:2008bj,Patel:2012pi,Blinov:2015sna,Niemi:2018asa,Niemi:2020hto}. }The \lq\lq h.c.\rq\rq is unnecessary for real $\phi$ and the $b_3$ term vanishes for the real triplet. Note that in the $Z_2$-symmetric limit, the neutral component of $\phi$ may acquire a vacuum expectation value, thereby spontaneously breaking fo the $Z_2$ symmetry. For gauge singlets, this situation results in the existence of cosmic domain walls, which can be problematic\cite{Zeldovich:1974uw,Kibble:1976sj,Kibble:1980mv}. For the real triplet, the constraints from the electroweak $\rho$ parameter\cite{Tanabashi:2018oca} imply that $\langle \phi^0\rangle / v \lsim 0.01$. For both the singlet and real triplet, the presence of a non-vanishing $a_1$ implies that the vev of $\phi$ cannot vanish. Given the $\rho$ parameter constraints, one immediately concludes that $a_1$ is at most of order a few GeV for the real triplet. For the gauge singlet case, no such constraints exist. In what follows, then, we will consider only scenarios with explicit $Z_2$ breaking and consider the consequences for collider phenomenology.

\vskip 0.25in

\noindent{\em Mixing angle}. When $\langle \phi^0\rangle\equiv x_0\not=0$, the interactions proportional to $a_1$ and $a_2$ will lead to mixing between the neutral components of $H$ and $\phi$, leading to two mass eigenstates $h_{1,2}$:
\bea
\label{eq:mixing}
h_1 & = h\cos\theta + \phi^0\sin\theta\\
\nonumber
h_2 & = - h\sin\theta + \phi^0\cos\theta
\eea
with the mixing angle given by
\be
\sin 2\theta = \frac{(a_1+2 a_2 x_0) v_0}{m_1^2-m_2^2}\ \ \ .
\label{eq:mixangle}
\ee
Note that the contribution from $a_2$ depends on its product with $x_0$.  The corresponding impact on the mixing angle can be vanishingly small for sufficiently small $x_0$, even for $a_2\sim\mathcal{O}(1)$. In contrast, the impact of the dimensionful parameter $a_1$ carries no such suppression, though in the limit of small $x_0$ one must also have a small $a_1$ due to the conditions for minimizing the scalar potential. 

As discussed above, at $T>0$, the cross-quartic operator proportional to $a_2$ can induce a barrier between the origin and the Higgs broken phase vev through thermal loops or between a high-temperature $\phi^0$ vacuum and the Higgs vev as a tree-level vev. Here, we concentrate on the role played by non-vanishing $a_1$. When $a_1<0$, Its presence also implies the existence of a tree-level barrier between the origin in field space and the vaccum located at $(\langle H^0\rangle, \langle \phi^0 \rangle) = (v_0/\sqrt{2}, x_0)$. A direct transition to this vacuum from the origin at $T>0$ -- as illustrated in Fig.~\ref{fig:scenarios}(c) -- will be first order owing to this tree-level barrier. 

At this stage, we possess no quantitative guidance for the values of $a_1$, $a_2 x_0$, and $\sin 2\theta$ other than from the electroweak $\rho$ parameter in the case of the real triplet. An additional consideration, however, may be drawn from the requirements for successful electroweak baryogenesis (EWBG) and/or the generation of observable gravitational waves (GW). Both rely on a first order EWPT that is sufficiently \lq\lq strong\rq\rq . The first order EWPT proceeds via bubble nucleation. In EWBG, the baryon asymmetry is generated when CP-violating interactions at the bubble walls induce a non-vanishing density of left-handed fermions in the unbroken phase (bubble exterior). The latter, in turn, biases the rapid  electroweak sphaleron (EWS) processes into generating a net  B$+$L asymmetry that diffuses inside the expanding bubbles. Preservation of this asymmetry in the Higgs phase  implies that the sphaleron processes inside the bubbles must be sufficiently quenched. The rate is given by (see Ref.~\cite{Quiros:1999jp} for a pedagogical discussion and references)
\be
\label{eq:Gsphaleron}
\Gamma_\mathrm{EWS}\sim A(T) \exp\left\{-E_\mathrm{sph}/T\right\}
\ee
where $E_\mathrm{sph}$ is the sphaleron energy and $A(T)$ is an in principle calculable prefactor. In the high-T effective theory, one has at leading order
\be
\frac{E_\mathrm{sph}}{T} \propto \frac{\bar{v} (T)}{T} \ \ \ ,
\ee
where $\bar{v}(T)$ is the Higgs vev in the LO high-temperature theory\footnote{The use of $\bar{v}(T)$ rather than the vev resulting from the minimization of the full finite-$T$ potential avoids any issues of gauge invariance. For a detailed discussion of this point, see Ref.~\cite{Patel:2011th}}. The relevant temperature in this case is the bubble nucleation temperature, $T_N$, which lies below $\tew$ (typically just below). As a rough estimate, the requirement that the initial bayon (B$+$L) asymmetry be preserved implies that
\be
\frac{\bar{v} (\tew)}{\tew} \gsim 1 \ \ \ .
\ee
Relating $\bar{v} (\tew)$ to the parameters of the scalar potential then yields, to a good approximation,
\be
\frac{\vert a_1\vert}{2\lambda\tew} \gsim 1\ \ \ .
\label{eq:a1req}
\ee
In the limit of negligible $a_2 x_0$, then, one has
\be
\vert \sin 2\theta \vert \approx \frac{\vert a_1\vert v}{\vert m_1^2-m_2^2\vert} \gsim \frac{2\lambda\tew v}{\vert m_1^2-m_2^2\vert}
\label{eq:mixanglereq}
\ee
To estimate the corresponding lower bound on the magnitude of $\vert \sin\theta\vert$ we take $m_1$ to be the observed Higgs-like boson mass and $m_2$ to be given conservatively by twice the upper bounds on mass range resulting from our earlier arguments\footnote{The results of model-dependent studies indicate that a strong first order transition is no longer viable for $m_2$ significantly above this scale
\cite{Profumo:2007wc,Profumo:2014opa,Kotwal:2016tex}
.}, or $m_2\approx 700$ GeV. We then obtain 
\be
\vert \sin\theta\vert \gsim 0.01\ \ \  .
\label{eq:mixanglebound}
\ee

Eq.~(\ref{eq:mixanglebound}) sets a scale for precision Higgs studies, although the foregoing arguments are not as air tight as those leading to the upper bound on mass scale. The presence of non-vanishing $a_2 x_0$ may lead to cancellations between the two terms in the numerator of Eq.~(\ref{eq:mixangle}), leading to smaller values of  $\vert \sin\theta\vert$, and the value of $\tew$ can be a factor of a few smaller than given in Eq.~(\ref{eq:TEWvalue}) due to additional contributions from the new scalars. Indeed, explicit studies\cite{Profumo:2014opa} indicate that the individual terms in the numerator of Eq.~(\ref{eq:mixangle}) can be larger in magnitude than implied by Eq.~(\ref{eq:a1req}), while leading to a small mixing angle. Nevertheless, these studies also indicate that -- assuming a flat prior for the choice of scalar potential parameters -- the typical magnitude of  $\vert \sin\theta\vert$ lies well above 0.01 for the vast majority of cases. Thus, it appears that the arguments leading to Eq.~(\ref{eq:mixanglebound}) do \mrm{indeed} yield a robust guide to the scale of precision needed to see the impact of a strong first order EWPT associated with explicit $Z_2$-breaking.

\vskip 0.25in

\noindent{\em Higgs boson self-coupling}. The Higgs boson self-coupling is a key parameter in the dynamics of EWSB. The presence of a non-vanishing $\sin\theta$ would imply a change in the strength of the Higgs boson triple self coupling\cite{Profumo:2007wc}
\be
\lambda_{hhh}\rightarrow \lambda_{111} = \lambda\cos\theta^3 +\frac{1}{4v}\left(a_1+2 a_2 x_0\right)\cos^2\theta\sin\theta + \cdots
\ee
where the $+\cdots$ indicate higher order terms in $\sin\theta$ that are negligible for purposes of our discussion. The magnitude of this change, relative to the SM prediction $\lambda_{hhh}=\lambda$ is given by 
\be
\left\vert \frac{\Delta\lambda}{\lambda}\right\vert\equiv \left\vert \frac{\lambda_{111} - \lambda_{hhh}}{\lambda_{hhh}}\right\vert =  \frac{\vert(a_1+2 a_2 x_0)\sin\theta\vert}{4\lambda v} +\cdots \ \ \ .
\ee
Consider now the small $x_0$ regime in which the $2 a_2 x_0$ term is negligible. In this case, one has
\be
\left\vert \frac{\Delta\lambda}{\lambda}\right\vert  \rightarrow \frac{\vert a_1\sin\theta\vert}{4\lambda v} \gsim 0.01 \times \frac{\tew}{2 v} \approx 0.003\ \ \ ,
\label{eq:deltalambound}
\ee
where we have used the inequalities in Eqs. (\ref{eq:mixanglebound}) and (\ref{eq:a1req}). The bound in Eq.~(\ref{eq:deltalambound}) is again not airtight but consistent with the result of explicit model studies \mrm{(see, {\em e.g.}, Refs.~\cite{Profumo:2014opa,Huang:2016cjm})}. As in the case of the mixing angle, these results indicate that considerably larger magnitudes for the shift in the triple self-coupling are favored. 

\vskip 0.25in

In both this and the previous section, we observe that the quantities $a_2$ and $x_0$ remain the least constrained by EWPT considerations, at least at this simple level of perturbative analysis. For the $Z_2$-symmetric case, the mass of the new scalar is fixed by $\tew$ and $a_2$ (and to a lesser extent $b_4$) once we require the presence of a first order EWPT. In the presence of explicit $Z_2$-breaking, the sign and magnitude of $a_2$ will determine whether the expected lower bounds $|\sin\theta|$ and $|\Delta\lambda/\lambda|$ are given by Eqs.~(\ref{eq:mixanglebound}) and (\ref{eq:deltalambound}), respectively, or whether cancellations imply their circumvention. At present, we have no generic probe of $x_0$ as an independent parameter, though detailed study of $\phi$ decays can provide indirect information. 

A first order EWPT in the $Z_2$ symmetric case requires $a_2>0$, implying a relative decrease in Higgs diphoton decay rate. In the presence explicit of $Z_2$-breaking, a positive value for $a_2$ -- implying a decrease in the di-photon decay rate -- would allow for cancellations in the quantity $a_1+2 a_2 x_0$ that governs the mixing angle and triple self-coupling, thereby allowing for one to circumvent the bounds in Eqs.~(\ref{eq:mixanglebound}) and (\ref{eq:deltalambound}). On the other hand, a negative value for this parameter -- implying an increase in the diphoton decay rate -- would preclude the possibility of such cancellations. Here again, an ultra-precise determination of the Higgs di-photon decay rate would determine the sign of $a_2$ and thereby potentially help solidify our expectations for the magnitude of $\sin\theta$.

\section{Collider Phenomenology: the LHC and Beyond}
\label{sec:collider}

The foregoing discussion provides concrete, benchmark mass and precision targets for present and prospective future colliders. One may now ask: What capabilities would be required to reach these benchmarks? Are these capabilities within the realm of the LHC or next generation colliders? In what follows, I provide simple estimates for the mass and precision reach of prospective future colliders as they bear on these benchmarks. One should bear in mind that these estimates are intended to be indicative of what may be possible in the future and that they do not constitute definitive studies. The latter, which require detailed simulations of signal and background events, detector capabilities, {\it etc.} go beyond the scope of the present work. 

I first consider the mass reach. If the new scalars are charged under SU(3$)_C$, then present LHC exclusion limits on various observables implies severe constraints for masses below one TeV (for a discussion, see, {\it e.g.}, Ref.~\cite{Katz:2014bha}). Consequently, I will focus on $\phi$ being an SU(3$)_C$ singlet.

\vskip 0.25in

\noindent{\em Electroweak pair production.}.
In the case of electroweak multiplets, scalars may be pair produced through electroweak Drell-Yan processes, such as $e^+e^-\to \phi^+\phi^-$ or $pp\to \phi^+\phi^0X$. In either case, the leading order (LO) partonic cross section for the process $f_1{\bar f_2}\to V^\ast\to\phi_1\phi_2$ mediated by a virtual gauge boson $V=\gamma$, $Z$, or $W^\pm$  with mass $M_V$ is 
\be
\label{eq:ewdy1}
{\hat\sigma} (f_1{\bar f_2}\to V^\ast\to  \phi_1\phi_2)  =  g_\phi^2 \times \mathcal{G}_V \times F_V(\shat, M_\phi)\ \ \ .
\ee
Here, 
\be
\mathcal{G}_V  = \left(\frac{g^4}{4\pi}\right)\, \left(\frac{g_V^2+g_A^2}{12}\right) v^{-2}\\
\ee
where $g$ is the gauge coupling;  $g_V$ ($g_A$) is the vector (axial vector) coupling of the parton pair $f_1{\bar f_2}$ to $V$; $g_\phi$ is the corresponding coupling to the $\phi_1\phi_2$ pair; $v=246$ GeV is the Higgs vacuum expectation value; and
\be
\label{eq:ewdy2}
F_V(\shat, M_\phi) = \left(\frac{v^2}{\shat}\right) \frac{1}{(1-M_V^2/\shat)^2}  \left(1-\frac{4 M_\phi^2}{\shat}\right)^{3/2}
\ee
with $\shat$ being the parton center of mass energy.
Here, we have not included the vector boson decay width $\Gamma_V$, though one could easily do so by replacing the $V$ propagator-squared by the appropriate Breit-Wigner formula. For $2M_\phi >> M_Z$ as implied by LEP limits\footnote{\mrm{The limits on charged scalar production can be inferred from the bounds on sleptons. For a discussion of the latter, see, {\em e.g.} Refs.~\cite{Lipniacka:2002sw,Ask:2003pg,Eckel:2014dza}. The most model-independent limits result from the search for a dilepton pair plus missing energy. For example, the lower bound on the right-handed selectron is 99.6 GeV for a selectron-neutralino mass splitting larger than 15 GeV. The left-handed selectron bounds are stronger. For a real electroweak multiplet with zero hypercharge, the corresponding production cross section is larger than for the left-handed sleptons, implying a bound above 100 GeV. For a mass splitting between the charged and neutral components of the multiplet smaller than 15 GeV, the corresponding LEP limits on charginos provide a benchmark. In this case, the lower bound lies above 90-100 GeV (see, {\em e.g.} Refs.~\cite{Lipniacka:2002sw,Ask:2003pg,Egana-Ugrinovic:2018roi}).    }}, the impact of including $\Gamma_V$ will not be appreciable. We have also normalized the function $F_V$ and prefactor $\mathcal{G}_V$ so that the former is dimensionless and the latter has the dimensions of a cross section. To set the scale, one has for a process mediated by a virtual $W$ boson $\mathcal{G}_W \approx 980$ fb. 

\begin{figure}[t]
\begin{center}
\includegraphics[width=9cm,height=7.5cm,angle=0]{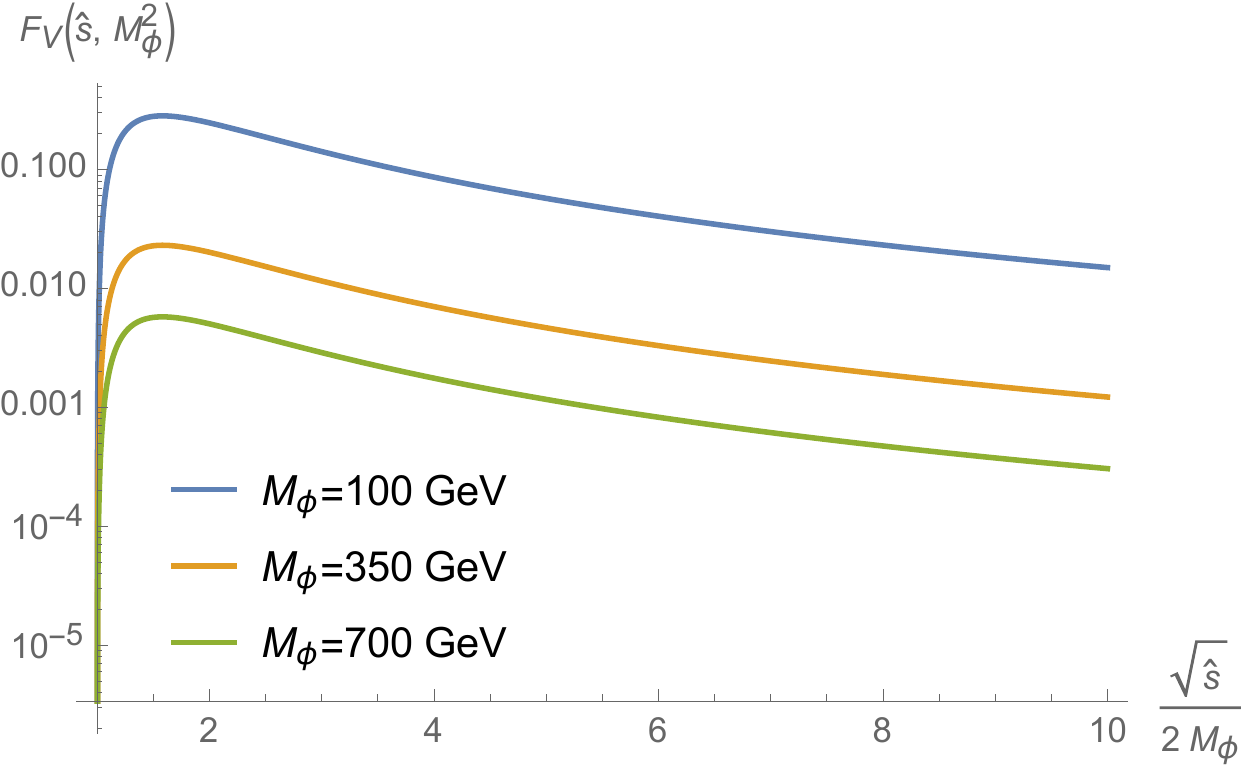}
\end{center}
\caption{The function $F_V({\hat s}, M_\phi^2)$ entering the partonic cross section for electroweak Drell-Yan pair production (\ref{eq:ewdy1},\ref{eq:ewdy2}).  }  \label{fig:epemfv}\end{figure}

Focusing first on prospective $e^+e^-$ colliders, we discuss three options under consideration: the International Linear Collider (ILC)\cite{Baer:2013cma}; a circular $e^+e^-$ collider as proposed for either the Circular Electron-Positron Collider (CEPC) in China\cite{CEPCStudyGroup:2018ghi} or the CERN Future Circular Collider (FCC) in the $ee$ mode\cite{Benedikt:2018qee}; and the Compact Linear Collider proposed for CERN\cite{Robson:2018zje,deBlas:2018mhx}. The center of mass energies $\sqrt{s}$ are set at specific values for these facilities. I take the following: $\sqrt{s} = 500$ GeV (ILC); 240 GeV (CEPC/FCC-$ee$); 340 GeV (FCC-$ee$); and 1.5 TeV and 3 TeV (CLIC), where the latter give the middle and highest value of the three center of mass energy options under study. It is worth noting that due to the fixed beam energies, the different facilities would have greatest sensitivity to $\phi$ pair production for different values of $M_\phi$.  To illustrate the peak sensitivities, we plot in Fig.~\ref{fig:epemfv} the function $F_Z(\shat, M_\phi)$ for representative values of $M_\phi$ in the EWPT target range, starting with $M_\phi=100$ GeV as a rough lower bound implied by LEP limits. 

Having scaled the parton center of mass (CM) energy by $2 M_\phi$, we observe a universal behavior, with a maximum occurring at $\sqrt{\shat}/2M_\phi\approx 1.7$ for all values of $M_\phi$ but with the magnitude of $F_Z$ dropping by about an order of magnitude for each representative choice of $M_\phi$. Thus, for a given $e^+e^-$ CM energy $E_\mathrm{CM}$, the maximal sensitivity will be for a scalar mass $\sim E_\mathrm{CM}/3.4$. To be concrete, the CLIC 1.5 TeV option would be best suited to $M_\phi\approx 440$ GeV, while a 500 GeV ILC would having maximum sensitivity to a mass roughly 150 GeV. Similarly, the FCC-$ee$ with $E_\mathrm{CM}=340$ GeV would be ideally suited to probing a 100 GeV new scalar. For $M_\phi$ near the upper end of our conservative EWPT-viable range, the optimal CM energy is roughly 2.4 TeV. The degradation in sensitivity by going to higher energy, such as the CLIC 3 TeV option, is modest. Note, however, that for a given beam energy, the cross section drops quickly with increasing $M_\phi$, going to zero as $M_\phi\to E_\mathrm{CM}/2$.

With this information in hand, it is straightforward to determine the number of produced $\phi$ pairs for a given $M_\phi$, $E_\mathrm{CM}$, and integrated luminosity. In Table \ref{tab:epem}, we give this information for each prospective collider, choosing $M_\phi$ in each case to give the maximum cross section. For purposes of illustration, we will assume the scalar multiplet is a real electroweak triplet and that the final state consists of a $\phi^+\phi^-$ pair. 
We take as projected design integrated luminosities as given in the fourth column of Table \ref{tab:epem}. The anticipated numbers of signal events are shown in the final column. 

In general, it is evident that even for new scalars at the upper end of the conservative EWPT mass range, the various $e^+e^-$ colliders will yield 10, 000 or more signal events. Given the clean environment for these colliders, observation of a signal should in principle be feasible. Obtaining concrete projections will require more detailed information about the expected signature, detector resolution, efficiency and other experimental details. For example, in the absence of $Z_2$-breaking interactions, the neutral component of $\phi$  may be stable. Electroweak radiative corrections will increase the mass of the components of charge $Q$ with resect to the neutral state by $M_Q- M_0 \approx Q^2\Delta M$, with $\Delta M = (166\pm 1)$ MeV\cite{Cirelli:2005uq}. The $\phi^\pm$ will thus decay to the $\phi^0$ plus a soft lepton pair or soft pion that is difficult to detect, yield a disappearing charged track (DCT) \cite{FileviezPerez:2008bj}. The detectability of the DCT will depend on the $\phi^\pm$ lifetime, detector resolution, and trigger. Assuming these issues are addressed, the upper limit $\phi^\pm$ mass reach will depend on the collider CM energy. 

\begin{table}[h]
\centering
\begin{tabular}{|c|c|c|c|c|}
\hline
\hline $E_\mathrm{CM}$(GeV) &  $M_\phi$ (GeV) &   $\hat{\sigma}$ (fb) & $\int dt\mathcal{L}$ (ab$^{-1}$) & $N\times 10^{-3}$ \\
\hline
\hline
340 & 100 &142 fb &  5  & 710  \\
\hline
500  & 100 & 94 fb &  2  & 188  \\
  & 150 & 63 fb &  2  & 126  \\
\hline
1500  & 150 & 13 fb &  2.5  & 32.5  \\
  & 440 & 7 fb &  2.5  & 17.5  \\
\hline
3000 & 440  & 3 fb &  5  & 15  \\
 & 700 & 2 fb &  5  & 10  \\
\hline
\hline
\end{tabular}
\caption{ Comparison of a circular $e^+e^-$ collider and two linear $e^+e^-$ options (ILC-500 and CLIC) for neutral current production of a $\phi^+\phi^-$ pair for representative choices of $M_\phi$. Final column contains the expected number of signal events ($N$) for the given cross section and integrated luminosity. } \label{tab:epem}
\end{table}

We now turn to the corresponding analysis for $pp$ collisions. In this case, while the beam energy is fixed, the parton CM energy is not. Instead, one must integrate over the parton distribution functions (pdfs), leading to the following expression for the cross section $\sigma(pp\to \phi_1\phi_2 X)$:
\be
\label{eq:ppxs}
\sigma(pp\to V^\ast\to \phi_1\phi_2 X) = \sum_{a,b}\ \int_{\shat_0}^\infty\ d\shat\ \left(\frac{d\mathcal{L}_{ab}}{d\shat}\right)\ {\hat\sigma}(ab\to V^\ast\to\phi_1\phi_2)\ \ \ ,
\ee
where the sum is over all partons $a$ and $b$ in the colliding protons, $\sqrt{\shat_0}=2 M_\phi$, and $d\mathcal{L}_{ab}/d\shat$
is the parton luminosity function constructed from the pdfs, suitably evolved to the energy scale of the partonic sub-process. We consider the charged current (CC) process $pp\to W^{+\ast}\to \phi^+\phi^0$ as the factor $\mathcal{G}_W$ is larger than the corresponding factors for the neutral current pair production . 

\begin{figure}[t]
\begin{center}
\includegraphics[width=9cm,height=7.5cm,angle=0]{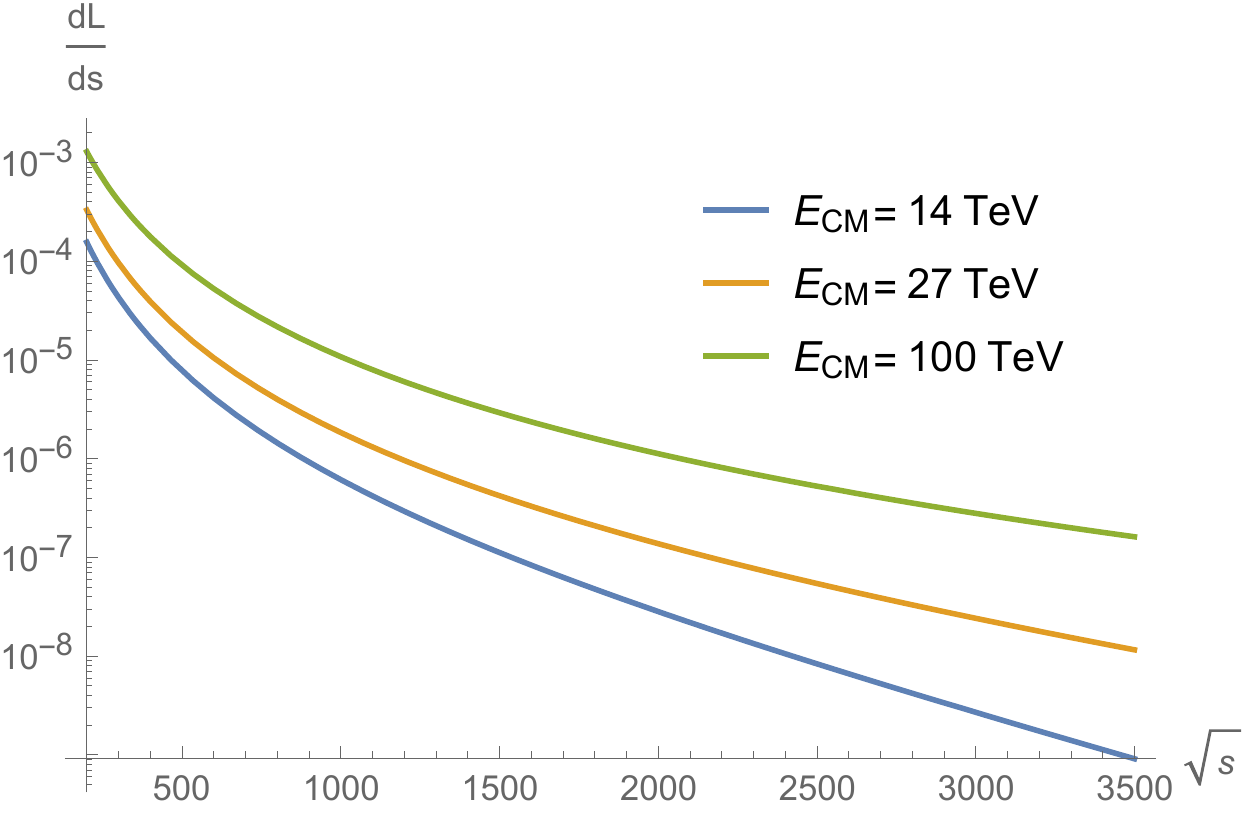}
\end{center}
\caption{Parton luminosities for charged current scalar pair production at different $pp$ collider CM energies.  }  \label{fig:cclumi}\end{figure}

For purposes of comparing different collider options, it is useful to plot $d\mathcal{L}_{ab}/d\shat$ for CC processes as a function of $\shat$ for three different CM energies: 14 TeV, 27 TeV, and 100 TeV, corresponding respectively to the HL-LHC\cite{ApollinariG.:2017ojx}, HE-LHC\cite{Cepeda:2019klc}, and either the FCC-hh\cite{Benedikt:2018csr} or SppC.  Recalling that for a given $M_\phi$ the optimal parton CM energy is $\sim 3.4 M_\phi$, we see that for a 700 GeV particle, a 100 TeV $pp$ collider will have roughly 60 times more signal events than the LHC, assuming the same integrated luminosity. Given the proposed FCC-hh integrated luminosity of 30 ab$^{-1}$, the total number of signal events would be 600 times greater than for the HL-LHC. To make this comparison more concrete, I provide in Table \ref{tab:ppDY} the cross sections and expected number of signal events for representative values of $M_\phi$, assuming the design integrated luminosities for the LHC, HE-LHC, and FCC-hh. Note that the results shown are based on LO cross sections, computed independently using the  parton luminosity functions obtained with CTEQ pdfs in the package ManeParse\cite{Clark:2016jgm} and directly using the pdf set cteql6\cite{Pumplin:2002vw}. The corresponding K-factor for the LHC with $\sqrt{s}=13$ TeV is the same as for slepton pair production, as both the scalars $\phi$ and sleptons carry only electroweak quantum numbers\footnote{The corresonding LO computation is consistent with the slepton pair production cross sections given in Ref.~\cite{Eckel:2014dza} after taking into account the difference in the real triplet and scalar doublet couplings to the $W$ boson}. The resulting values are modest: $K=1.18$ at NLO \cite{Fuks:2013lya}, with small corrections of order a percent arising from next-to-leading logarithm (NNL) and next-to-next-to-leading logarithmic (NNLL) resummations matched to approximate next-to-next-to-leading order (aNNLO) QCD corrections\cite{Fiaschi:2019zgh}. To my knowledge, the corresponding computations do not exist for the HL-LHC or a 100 TeV $pp$ collider, so for purposes of comparison among the different collider options I do not apply a $K$-factor correction for the HL-LHC results. In this context, detailed phenomenological studies of the real triplet phenomenology at the LHC and a 100 TeV $pp$ collider have appeared in Refs.~\cite{Bell:2020gug,Chiang:2020rcv}.

\vskip 0.25in

\noindent{\em Singlet-like scalar production}. For SM gauge singlets, DY pair production rates will be highly suppressed by four powers of the small singlet-doublet mixing angle. On the other hand, production of one or more singlet-like scalars may occur at appreciable rates via the following mechanisms: 
\begin{itemize}
\item[(i)] \textit{Single scalar production through mixing}. The production cross sections for production of one singlet-like scalar having mass  $m_2$ [see Eq.~(\ref{eq:mixing})] will go as $\sin^2\theta$ times the cross section for production of a single purely SM Higgs boson with  mass $m_2$. For $m_2< 2 m_1$, the $h_2$ decay branching ratios will be identical to those of a SM Higgs boson of the same mass. For heavier $m_2$, the decay $h_2\to h_1 h_1$ is kinematically allowed, and the corresponding branching ratios to the SM Higgs decay final states and the di-Higgs state will depend in detail on the model parameters. 
\item[(ii)] \textit{Pair production through the Higgs portal}. In the limit of vanishing mixing angle, the $h_1 h_2 h_2$ coupling remains non-zero and is proportional to $a_2$. Thus, one may consider the process $pp\to h_1^{(\ast)}\to h_2 h_2$ in the absence of mixing (see, {\it e.g.}, Ref.~\cite{Chen:2017qcz}) and $pp\to h_2^{(\ast)}\to h_2 h_1$ for non-zero $\sin\theta$. In the former instance, for $m_2< m_1/2$, the intermediate Higgs-like scalar may be on-shell, leading to \lq\lq exotic Higgs decay" modes. 
\end{itemize}

In what follows, I consider the simpler case of single scalar production (i) and comment briefly on the other cases below. Our focus here will suffice to illustrate the potential mass reach of the LHC and prospective future colliders. To that end, we study the associated production mechanism $e^+e^-\to Z^\ast\to Z h_2$ and the gluon-gluon fusion ($gg$F) production mechanism that gives the largest \lq\lq heavy Higgs" production cross section in $pp$ collisions. 

\vskip 0.25in
\noindent {\em Associated heavy Higgs production in $e^+e^-$ annihilation}. The LO partonic associated production cross section for a purely SM Higgs boson is given by \be
\label{eq:zh}
{\hat\sigma}(f{\bar f}\to Zh) = \frac{2\pi\alpha^2(g_V^2+g_A^2)}{48 N_C (s_W c_W)^4} \frac{2k}{\sqrt{s}} \frac{k^2+3 M_Z^2}{(s-M_Z^2)^2}\ \ \ 
\ee
where $k$ is the Higgs boson momentum in the partonic CM frame and $s_W$ ($c_W$) is the sine (cosine) of the weak mixing angle. In the presence of $h$-$\phi^0$ mixing, the cross section for associated production of the $h_1$ state will be the same as in Eq.~(\ref{eq:zh}) but multiplied by $\cos^2\theta$. For $|\sin\theta|$ given by the lower bound in Eq.~(\ref{eq:mixanglebound}), the resulting decrease in the SM associated production cross section will be far too small to be observable in the proposed leptonic Higgs factories. 

In principle, a more promising avenue could be direct production of the state $h_2$ using associated production with a higher-energy lepton collider. In practice, it appears difficult to achieve sufficient statistics with any of the proposed lepton colliders. To illustrate, we give in Table \ref{tab:eesinglet} the cross sections and corresponding expected number of events for different $e^+e^-$ CM energies and integrated luminosities for representative values of $M_\phi$ and a $|\sin\phi| = 0.01$. Except for the lightest values of $M_\phi$ at the lower CM energies, the cross section is too small to yield any signal. On the other hand, one may use the values for $\sigma$ and $\int dt\mathcal{L}$ to determine the minimum $|\sin\theta|$ that one might probe for a given $M_\phi$. In short,  a complete probe of a $H^\dag H\phi$-induced strong first order EWPT using associated production does not appear to be possible with any of the currently envisioned new lepton colliders. However, a significant portion of the relevant parameter space would still be experimentally accessible.

\begin{figure}[t]
\begin{center}
\includegraphics[width=9cm,height=7.5cm,angle=0]{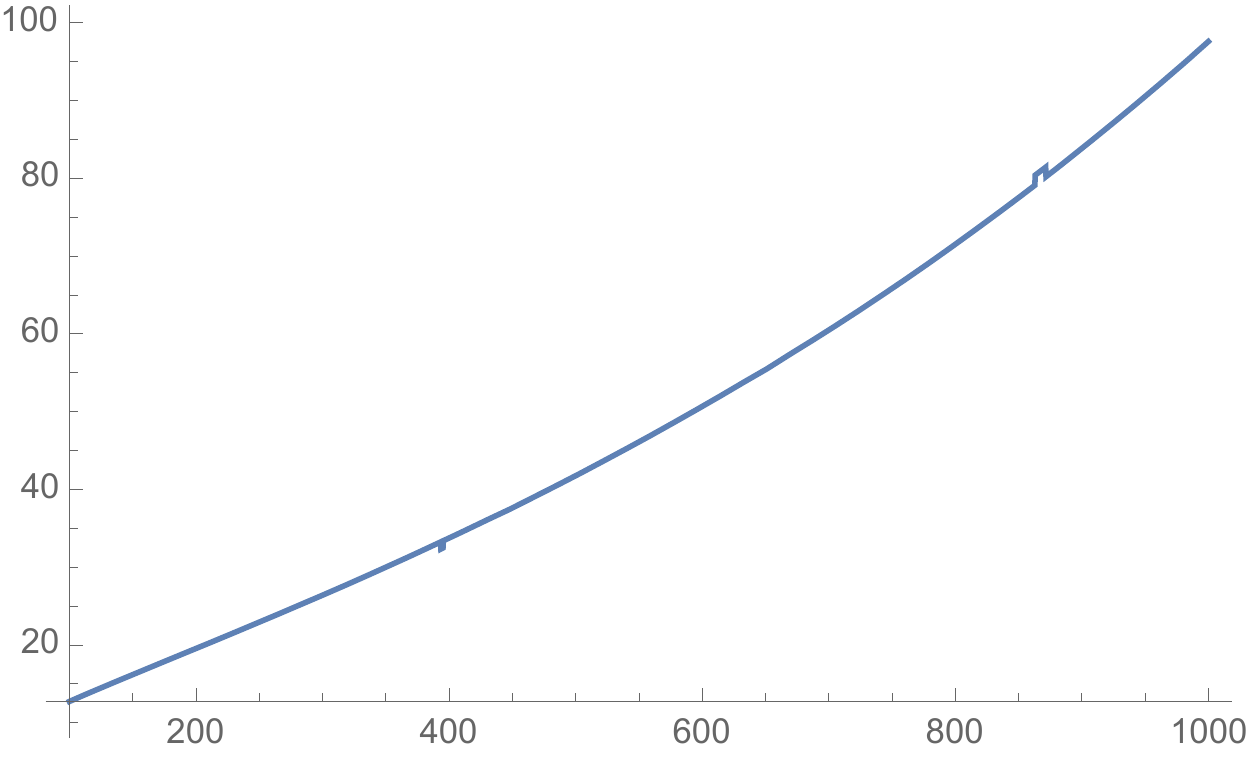}
\end{center}
\caption{Ratio of the gluon luminosity in $pp$ collisions at CM energies 100 TeV to  that at 14 TeV as a function of parton CM energy $\sqrt{\hat s}$. }  \label{fig:gglumi}\end{figure}

\vskip 0.25in
\noindent{\em Gluon fusion heavy Higgs production in $pp$ collisions}. The cross sections for $gg$F production for a heavy SM Higgs boson for $\sqrt{s}=14$ TeV have been tabulated by the LHC Higgs Cross Section Working Group. To obtain the corresponding $h_2$ production cross section, one simply scales the SM cross sections by $\sin^2\theta$. The corresponding cross sections for higher $pp$ CM energies requires use of Eq.~(\ref{eq:ppxs}). Recall that the  parton luminosity as a function of $\shat$  varies with $pp$ CM energy, as previously illustrated for the CC DY process in Fig.~\ref{fig:cclumi}. 
To gain a rough idea of the impact of the difference in parton luminosity, we plot in Fig.~\ref{fig:gglumi} the ratio of parton luminosities for the $gg$F process at 14 and 100 TeV. As an illustration, for threshold production of an on-shell $h_2$ with $m_2=700$ GeV, the parton luminosity at a 100 TeV collider is roughly sixty times larger than at the LHC. The corresponding gain in $\sigma(pp\to h X)$ will be larger due to the integral in Eq.~(\ref{eq:ppxs}). As an illustration, we give in Table \ref{tab:ppsinglet} the production cross sections for representative masses and mixing angles given in Refs.~\cite{Kotwal:2016tex,Huang:2017jws,Li:2019tfd} after rescaling to the benchmark lower value of $|\sin\theta|$ given in Eq.~(\ref{eq:mixanglebound}). At the upper end of the target mass range, one would expect at most several hundred signal events at the HL-LHC, implying that discovery would be challenging at best. At the higher energy and design luminosity of a 100 TeV $pp$ collider, on the other hand, one would anticipate several hundred thousand events. In referring to the values in  Table \ref{tab:ppsinglet}, one should bear in mind that the values of $|\sin\theta|$ obtained in Refs.~\cite{Kotwal:2016tex,Huang:2017jws,Li:2019tfd} are considerably larger than 0.01. The results in these studies were obtained by scanning over the parameters of the potential in Eqs.~(\ref{eq:v0phi},\ref{eq:v0hphi},\ref{eq:z2break}), and requiring that the first order EWPT completes ({\it e.g.}, a sufficiently large tunneling rate) and that the baryon number preservation criterion be satisfied. Hence, the benchmarks given in Table \ref{tab:ppsinglet} appear to be quite conservative.

\begin{table}[h]
\centering
\begin{tabular}{|c|c|c|c|c|}
\hline
\hline $E_\mathrm{CM}$(TeV) &  $M_\phi$ (GeV) &   ${\sigma}$ (fb) & $\int dt\mathcal{L}$ (ab$^{-1}$) & $N\times 10^{-3}$ \\
\hline
\hline
14 & 415 & $7.7$ &  3  & 23  \\
 & 714 & $ 0.63$ &  3  & 1.9  \\
 \hline
27 & 415& 26  &  30  & 720  \\
  & 714 & 3 &  30  & 90  \\
\hline
100 & 415& 183 &  30  & 5490  \\
  & 714 & 29  &  30  & 870  \\
\hline
\hline
\hline
\end{tabular}
\caption{Comparison of the LHC, HE-LHC and 100 TeV pp sensitivities to $\phi^+\phi^0$ electroweak Drell-Yan production  for representative choices of $M_\phi$. Note that  $K$-factors have not been applied, as discussed in the text. }\label{tab:ppDY}
\end{table}

\begin{table}[h]
\centering
\begin{tabular}{|c|c|c|c|c|c|}
\hline
\hline $E_\mathrm{CM}$(TeV) &  $M_\phi$ (GeV) & $|\sin\theta|$ &   ${\sigma}$ (fb) & $\int dt\mathcal{L}$ (ab$^{-1}$) & $N$ \\
\hline
\hline
340 & 150 & 0.01 & 0.01  &  5  & 50  \\
\hline
500 & 150 & 0.01 & 0.005  &  2  & 10  \\
& 240 & 0.01 & 0.003 & 2 & 6 \\
\hline
1500 & 150 & 0.01 &  $5\times 10^{-4}$   &  2.5  & $ 1$  \\
  & 400 & 0.01 & $4\times 10^{-4}$   & 2.5 & $ 1$\\
  & 700  & 0.01 &  $2\times 10^{-4}$  & 2.5 &  $<1$ \\
\hline 
\hline
3000 & 150 & 0.01 &  $1 \times 10^{-4}$   & 5  & $<1$  \\
  & 400  & 0.01 & $1\times 10^{-4}$   & 5 & $<1$\\
  & 700 & 0.01 &  $1\times 10^{-4}$  & 5 & $<1$ \\ 
\hline

\end{tabular}
\caption{ Single heavy Higgs production via associated production at prospective $e^+e^-$ colliders. } \label{tab:eesinglet}
\end{table}


\begin{table}[h]
\centering
\begin{tabular}{|c|c|c|c|c|c|}
\hline
\hline $E_\mathrm{CM}$(TeV) &  $M_\phi$ (GeV) & $|\sin\theta|$ &   ${\sigma}$ (fb) & $\int dt\mathcal{L}$ (ab$^{-1}$) & $N\times 10^{-3}$ \\
\hline
\hline
14 & 415 & 0.01 &  $1$   &  3 & 3 \\
& 714 & 0.01 & $0.1$   & 3 &  0.3 \\
\hline
100 & 415 & 0.01 & 59  &  30  & 1770  \\
& 714 & 0.01 & 12  & 30 & 360\\
\hline
\hline
\end{tabular}
\caption{ Comparison of the LHC and 100 TeV pp sensitivities to $\phi^0$ production via the gluon fusion process, rescaling the cross sections given in Refs.~\cite{Kotwal:2016tex,Huang:2017jws,Li:2019tfd} by the minimum 
$|\sin\theta|$ of Eq.~(\ref{eq:mixanglebound}) for representative choices of $M_\phi$.} \label{tab:ppsinglet}
\end{table}


\section{Other Considerations}
\label{sec:other}

The foregoing discussion illustrates \mrm{quantitatively} how dynamics that modify the thermal history of EWSB and lead to a first order EWPT cannot involve new particles that are arbitrarily heavy or interact too feebly with the SM Higgs boson. The possible signatures for collider probes generally lie well within the reach of the LHC and/or prospective future colliders under consideration. The results of detailed studies within specific models\cite{Espinosa:1993bs,Benson:1993qx,Choi:1993cv,Vergara:1996ub,Ham:2004cf,Ahriche:2007jp,Profumo:2007wc,Noble:2007kk,Espinosa:2007qk,Espinosa:2008kw,Barger:2007im,Ashoorioon:2009nf,Das:2009ue,Espinosa:2011ax,Cline:2012hg,Chung:2012vg,Barger:2011vm,Huang:2012wn,Damgaard:2013kva,Fairbairn:2013uta,No:2013wsa,Profumo:2014opa,Craig:2014lda,Curtin:2014jma,Chen:2014ask,Katz:2014bha,Kozaczuk:2015owa,Kanemura:2015fra,Damgaard:2015con,Huang:2015tdv,Kanemura:2016lkz,Kotwal:2016tex,Brauner:2016fla,Huang:2017jws,Chen:2017qcz,Beniwal:2017eik,Cline:2017qpe,Kurup:2017dzf,Alves:2018jsw,Li:2019tfd,Gould:2019qek,Kozaczuk:2019pet,Carena:2019une,Branco:1998yk,Barger:2008jx,Jiang:2015cwa,Chiang:2017nmu,FileviezPerez:2008bj,Chowdhury:2011ga,Patel:2012pi,Blinov:2015sna,Niemi:2018asa,Chao:2018xwz,Bell:2020gug,Chiang:2020rcv,Niemi:2020hto,Carena:1996wj,Delepine:1996vn,Cline:1996cr,Laine:1998qk,Carena:2008vj,Cohen:2012zza,Laine:2012jy,Curtin:2012aa,Carena:2012np,Katz:2015uja,Pietroni:1992in,Davies:1995un,Huber:2000mg,Ham:2004nv,Ham:2004pd,Menon:2004wv,Funakubo:2005pu,Huber:2006ma,Chung:2010cd,Kozaczuk:2014kva,Huang:2014ifa,Huang:2016cjm} are broadly consistent with these simple, more general arguments. In fact, the requirements on mass and precision reach obtained in model realizations are generally more  optimistic than those appearing above. Thus, we can be fairly confident in our primary conclusion that $\tew$ sets a concrete, well-defined scale for new dynamics that collider studies may, in principle, probe exhaustively. 

That being said, there remain a few other general considerations that one should address on this topic. 
\begin{itemize}
\item The foregoing arguments rely on the various patterns of symmetry breaking illustrated in Fig.~\ref{fig:scenarios}, driven by thermal loops involving the new degrees of freedom and/or tree-level barriers in the tree-level scalar potential at the renormalizable level. 
The presence of higher dimensional operators can play a role analogous to the tree-level barriers discussed above if the associated mass scale is not too heavy with respect to $\tew$. 
\item It is conceivable that the new particles associated with a first order EWPT are relatively light compared to $\tew$. It is natural, then, to ask about the collider reach for both direct and indirect searches.
\item The value of $\tew$ itself may change in the presence of new interactions, and one may wonder about the corresponding impact on the mass and precision targets discussed above. In particular, contributions from  loops at either $T>0$ or $T=0$ can lower the transition temperature under certain conditions. These changes in $\tew$ motivate, in part, the choice of a somewhat larger upper bound on the $M_\phi$ mass range compared to the values $\sim 360-375$ obtained from the simple arguments given above.
\item \mrm{ The discussion in this work has relied heavily -- but not exclusively -- on thermodynamic properties of the field theory at $T>0$. In the case of theories without an explicit 
$Z_2$-symmetry, however, the benchmark lower bound on the magnitude of the mixing angle obtained in subsection \ref{sub:z2break} relies on one dynamical property, namely the electroweak sphaleron rate. An equally important dynamical quantity is the bubble nucleation rate, $\Gamma_N$.
 In order for a first order EWPT to actually occur,  $\Gamma_N$ must be sufficiently large. Imposing this requirement typically imposes further restrictions on the parameter space compared to what is allowed by purely static, thermodynamic properties. Thus, the mass and precision bounds derived above are likely to be conservative. I elaborate further on this point below.}
\end{itemize}

In what follows, I comment briefly on each of these points.

\subsection{Non-renormalizable Interactions}
\label{sec:non-renorm}
The lowest-dimension non-renormalizable, gauge-invariant operators that contain only Higgs boson fields enter the Lagrangian at $d=6$.  Following Ref.~\cite{Grojean:2004xa}, consider the corresponding Higgs potential of the form
\be
{\tilde V}_0(H) = \lambda\left(H^\dag H - \frac{v^2}{2}\right)^2+\frac{1}{\Lambda^2}\left(H^\dag H- \frac{v^2}{2}\right)^3\ \ \ ,
\label{eq:vhd6}
\ee
where the notation ${\tilde V}_0$ indicates that the leading order scalar potential is distinct from the potential in Eq.~(\ref{eq:v0H}). In both cases, the potential minimum occurs at $\langle H^0\rangle = v/\sqrt{2}$ and the square of the Higgs boson mass is $m_h^2=2\lambda v^2$. 

Writing Eq.~(\ref{eq:vhd6}) in terms of the field $h$ gives
\be
{\tilde V}_0(h) = {\tilde V}_0 - \frac{{\tilde\mu}^2}{2} h^2 + \frac{\tilde\lambda}{4} h^4 + \frac{1}{8\Lambda^2} h^6
\ee
where
\be
{\tilde\mu}^2 = \left[\lambda-\frac{3 v^2}{4\Lambda^2}\right]v^2 \qquad , \qquad {\tilde\lambda} = \lambda-\frac{3v^2}{2\Lambda^2}\ \ \ .
\ee
For $\Lambda^2 < 3 v^2/\lambda = 3 v^4/m_h^2$, one has ${\tilde\lambda} < 0$. The presence of the negative quartic term corresponds to a barrier between the symmetric and broken phases at $T=0$. Given the measured value of $m_h$, one then requires the mass scale $\Lambda$ to be less than $\sim 840$ GeV.  The authors of Ref.~\cite{Grojean:2004xa} find that for the EWPT transition to be strongly first order, the upper bound on $\Lambda$ is reduced by roughly 5\% for $m_h=125$ GeV. These results are consistent with what has been obtained in similar work by the authors of Ref.~\cite{Huang:2015izx} (see also Ref~\cite{Grinstein:2008qi}).

\mrm{Note that in this scenario, the thermal history of EWSB corresponds to Fig.~\ref{fig:scenarios}(a), but with a tree-level rather than thermal-loop induced barrier. For sufficiently high $T$, thermal loop corrections change the sign of the quadratic term in the potential, restoring the EW symmetry in this regime. Below, I discuss the relationship between the transition temperature and $\tew$. An implicit assumption in this context is that  either (a) whatever heavy fields have been integrated out so as to yield the $(H^\dag H)^3$ term in  Eq.~(\ref{eq:vhd6}) do not obtain vacuum expectations values at an intermediate temperatures or (b) the Higgs is pseudo-Goldstone boson of a theory possibly involving strong dynamics\cite{Grinstein:2008qi}. Interestingly, the upper bound of $\Lambda\lsim 840$ GeV is commensurate with the conservative upper bound I obtain with explicit retention of new scalars in the theory at non-zero $T$.}

The experimental signatures of the $d=6$ interaction in the potential would be modifications of the Higgs boson properties. In particular, the authors of Ref.~\cite{Cao:2017oez} noted that the corresponding loop-induced change in the associated production process $e^+e^-\to Zh$ at $\sqrt{s}=250$ GeV would range from $\sim 1$ to $\sim 2.5\%$ for a values of $\Lambda$ leading to a strong first order EWPT. An effect of this magnitude would be well within reach of the Higgs factories presently under consideration.

\subsection{New Light Particles}
\label{sec:light}
For the case of $\phi$ being a SM gauge singlet, it was observed in Ref.~\cite{Profumo:2007wc} that a strong first order EWPT may arise even when $M_\phi$ may be substantially below $m_h$.  For $m_h> 2 M_\phi$, the $h$-$\phi$ interaction will then lead to new Higgs boson decay modes. For a potential having a $Z_2$ symmetry, these decays will be unobservable, as $\phi$ is stable and will leave no traces in the detector. In the presence of a broken $Z_2$ (either explicitly or spontaneously), the $\phi$-$h$ mixing will enable the $\phi$ to decay via all kinematically-allowed SM Higgs boson decay channels. If the latter are sufficiently prompt, one may search for these \lq\lq exotic" Higgs decay modes. According to the initial analysis of Ref.~\cite{Profumo:2007wc}, the corresponding exotic decay branching ratios could be signficant.

Two recent studies have analyzed this regime in detail\cite{Carena:2019une,Kozaczuk:2019pet}. The authors of Ref.~\cite{Kozaczuk:2019pet} showed that there exists a lower bound on the exotic decay branching ratio as a function of $M_\phi$ for choices of parameters yielding a strong, first order EWPT. This study considered the real singlet extension for two cases: (a) explicit $Z_2$-breaking and (b) $Z_2$-symmetric. For $M_\phi\gsim 10$ GeV, the a combination of the LHC and prospective future lepton colliders appear to have the sensitivity needed to probe these scenarios. The work of Ref.~\cite{Carena:2019une} considered spontaneous $Z_2$-breaking\footnote{The problematic domain walls may, in principle, be avoided through the inclusion of higher dimension operators.} and analyzed the reach of both collider studies and gravitational wave probes, indicating that at least some portion of the relevant parameter space would be experimentally accessible.

\subsection{The Electroweak Temperature Revisited}
\label{sec:tewrev}

Here, I consider three possible ways in which new interactions may modify $\tew$ from the SM value of roughly 140 GeV: (a) $T>0$ loops involving the new particles; (b) $T=0$ loops, encoded in the Coleman-Weinberg potential; and (c) non-renormalizable interactions. In each case, it is possible that $\tew$ can be lower than in the SM, leading to a somewhat larger value of the allowed $M_\phi$. 

\subsubsection{$T>0$ Loops and $\tew$}
\label{sec:tg0loops}
The $H^\dag H\phi^\dag\phi$ interaction in Eq.~(\ref{eq:v0hphi}) will generate new thermal contributions to the finite temperature effective potential. For $\phi$ being a real scalar singlet, the impact corresponds to adding $a_2$ to the denominator of Eq.~(\ref{eq:T0sq}), {\it viz}, 
\be
2 y_t^2+\cdots \rightarrow 2y_t^2+a_2+\cdots \ \ \ .
\ee 
For $a_2 = 5$, $\tew$ will be lowered to $\approx 90$ GeV. The corresponding increase in the value of $M_\phi(T=0)$ for either the one-step or two-step scenario will be relatively modest. For EWGB, on the other hand, the lower temperature will enable more effective baryon number preservation  via suppression of the broken phase sphaleron rate [see Eq.~(\ref{eq:Gsphaleron})].

\subsubsection{$T=0$ Loops: Coleman-Weinberg}
\label{sec:cw}
The authors of Refs.~\cite{Huang:2014ifa,Harman:2015gif} showed that for scenarios in which a first order EWPT occurs, the impact of $T=0$ loops involving new particles can lower the critical temperature, thereby enhancing the degree of baryon number preservation and the magnitude of the gravitational wave signal. It is less clear whether this effect alone could induce a first order EWPT without the additional contributions to the barrier as discussed above. The change in the transition temperature arises from a reduction in the difference between the broken and unbroken phase vacuum energies:
\be
\Delta V \equiv V(\varphi=0) - V(\varphi=v) \ \ \ .
\ee
At tree-level in the SM, this difference is given by
\be
\Delta V_0  = -\frac{\lambda v^4}{4} \ \ \
\label{eq:vac0}
\ee
The one-loop contribution can be obtained from the Coleman-Weinberg potential
\be
V_\mathrm{CW} = \sum_k \frac{(-1)^{2 s_k} }{64\pi^2} [M_k^2(\varphi)]^2 \left\{ \ln\frac{M_k^2(\varphi)}{\mu^2}-\frac{3}{2}\right\}
\label{eq:vcw}
\ee
where the sum is over all particles of spin $s_k$ that interact with the field $\phi$; $M_k(\varphi)$ are the corresponding field-dependent masses; and $\mu$ is the renormalization scale\footnote{In an unfortunate clash of notation, the symbol $\mu$ in the present discussion does not refer to the Higgs mass parameter appearing in the tree-level SM.}. Note that one may eliminate the $\mu$-dependence of the full effective potential by replacing the tree-level couplings and mass parameters by their $\mu$-dependent running values, with the latter determined by the appropriate renormalization group (RG) equation. By a suitable choice of $\mu$, one may in principle eliminate any explicit large logarithms from Eq.~(\ref{eq:vcw}) and re-sum them via the RG evolution. In particular, the vacuum energy difference $\Delta V_0$ will inherit these re-summed logarithms via the running Higgs quartic self-coupling, $\lambda(\mu)$. \footnote{One may similarly move the $-3/2$ constant term into Eq.~(\ref{eq:vcw}) by making the replacement 
$\mu^2\to \mu^2 \exp(-3/2)$ though there is no {\it a priori} reason to do so. }

The resulting vacuum energy difference will then be given by 
\bea
\nonumber
\Delta V  \,\, & = & \,\,  -\frac{\lambda(\mu) v^4}{4} + \sum_k \frac{(-1)^{2 s_k} }{64\pi^2}\Biggl\{
\frac{3}{2} \left( [M_k^2(\phi)]^2-[M_k^2(0)]^2\right)\\
&&+ \left([M_k^2(0)]^2 \ln\frac{M_k^2(0)}{\mu^2} - [M_k^2(\varphi)]^2 \ln\frac{M_k^2(\varphi)}{\mu^2}\right)\Biggr\}\ \ \ .
\eea
Depending on the initial conditions for the RGE, $\Delta V$ may be smaller than $\Delta V_0$, corresponding to a vacuum energy \lq\lq uplift" due to the $T=0$ loops. Explicit, model-dependent studies performed in the next-to-minimal supersymmetric Standard Model\cite{Huang:2014ifa} and Two Higgs Doublet Model\cite{Dorsch:2017nza} indicate that such a vacuum energy uplift may, indeed, occur. Na\"ively, it is straightforward to see how this situation may arise. If $\mu\sim M_k(0)\sim M_k(\varphi)$, so that the logarithmic contributions are negligible; and if $M_k^2(\varphi) > M_k^2(0)$, then the explicit loop contribution to $\Delta V$ will be positive.  In this case, the value of $\tew$ will be lower than given in Eq.~(\ref{eq:T0sq}) since it takes less thermal energy to heat the vacuum and restore the symmetry. \mrm{This effect may explain, at least in part, the results obtained in a variety of 2HDM studies\cite{Turok:1991uc,Davies:1994id,Hammerschmitt:1994fn,Cline:1996mga,Fromme:2006cm,Cline:2011mm,Dorsch:2013wja,Dorsch:2014qja,Harman:2015gif,Basler:2016obg,Dorsch:2017nza,Bernon:2017jgv,Andersen:2017ika,Kainulainen:2019kyp}.} On the other hand, if these assumptions do not apply, then the presence of BSM interactions may lower the broken phase vacuum energy. 

\subsubsection{Higher Dimension Operators and the Transition Temperature}
\label{sec:d6Tc}

For the potential of Eq.~(\ref{eq:vhd6}) determination of the critical temperature requires inclusion of the $T>0$ loops, which yield a one-loop contribution 
\be
\Delta{\tilde V}(h,T) = \frac{c}{2} T^2 h^2 + \cdots
\ee
where\cite{Grojean:2004xa}
\be
c= \frac{1}{8}\left[2 y_t^2+\frac{3}{2} g^2 +\frac{1}{2} g^{\prime\, 2} + 2\left(\frac{m_h^2}{v^2} -3\frac{v^2}{\Lambda^2}\right)\right]\ \ \ .
\label{eq:d6T}
\ee
Recalling that $m_h^2=2\lambda v^2$, one observes that the first four terms inside the square brackets in Eq.~(\ref{eq:d6T}) are the same as in the SM. The contribution from the $d=6$ operator coefficient will reduce the value of $c$ from its SM value. The critical temperature for the first order EWPT is then given by
\be
T_0^2 = \left[8\lambda\left(\lambda+\frac{v^2}{\Lambda^2}\right) -6\left(\frac{v}{\Lambda}\right)^4+\mathrm{loops}\right]
\left[2 y_t^2+\frac{3}{2} g^2 +\frac{1}{2} g^{\prime\, 2} + 4\lambda -6\left(\frac{v^2}{\Lambda^2}\right)\right]^{-1}
\frac{\Lambda^2}{4} \ \ \ .
\label{eq:T0sqd6}
\ee
This expression is analogous to the one given in Eq.~(\ref{eq:T0sq}), which defines $\tew$ in the SM, but differs in several crucial ways: (i) the overall scale is now set by $\Lambda/2$ rather than $v$ [the last factor in Eq.~(\ref{eq:T0sqd6})]; (ii) the dependence on $\lambda$ in the numerator is different; and  (iii) the contributions of $\mathcal{O}(v^2/\Lambda^2)$ modify the first two factors. From Eq.~(\ref{eq:T0sqd6}) we find that $T_0$ increases monotonically from 34 GeV to 108 GeV as $\Lambda$ increase from 500 to 840 GeV. 

\mrm{
\subsection{Nucleation Rate}
The nucleation rate per unit volume can be expressed as\cite{Linde:1977mm,Linde:1980tt,Linde:1981zj}
\be
\gamma_N \equiv \Gamma_N/(\mathrm{Vol}) \sim A(T) \exp{-S_3/T}
\ee
where $S_3$ is the three-dimension Euclidean action and the prefactor $A(T) \sim \mathcal{O}(1) \times T^4$ (for a pedagogical discussion, see, {\em e.g.} Ref.~\cite{Quiros:1999jp}). In order that a first order EWPT completes, $\gamma_N$ must be larger 	than the expansion rate of the universe. Typically, this requirement implies that $S_3/T\lesssim \mathcal{O}(130-140)$ \cite{Quiros:1999jp}. A choice of model parameters leading to a larger value may be consistent with the static, thermodynamic requirements discussed above (necessary conditions) but would not be sufficient for the occurrence of an actual transition. For this reason, the simplified model parameter space I have used in the foregoing discussion is likely to be overly generous. Nonetheless, collider searches that probe this larger parameter space and yield a null result would still constrain the EWPT-viable parameter space. On the other hand, the observation of a collider signal would not, by itself, guarantee that a first order EWPT occurred. 

In principle, the observation (or lack thereof) of a primordial gravitational wave (GW) signal would provide an additional test. In practice, the next generation of GW probes, such as the LISA mission\cite{Audley:2017drz}, may not yet have the requisite sensitivity. As an illustrative study, the authors of Ref.~\cite{Gould:2019qek} find that when the new scalar $\phi$ is sufficiently heavy to be integrated out of the three-dimensional, finite temperature effective field theory, one may expect bubble nucleation and GW production to occur but with a signal too weak to be observable with LISA. Thus, confirmation that a first order EWPT-consistent collider signature corresponds to an actual transition will require either a following-generation GW probe and/or theoretical input.

On the theory side, obtaining robust computations of $\gamma_N$ is an open problem. The standard approach in many phenomenology papers is to obtain the bounce solution\cite{Coleman:1977py,Callan:1977pt,Coleman:1980aw} for $\phi(0,{\vec x})$ from the equation of motion involving $V(h,\varphi, T)$ and compute the resulting bounce action to derive $S_3$. A recent advance for the multi-field problem has been the development of the {\em CosmoTransitions} code\cite{Wainwright:2011kj}. The results of numerical studies using these approaches -- particularly for the multi-field case --  is difficult to recast into the simple parametric form I have used in obtaining the mass and mixing angle bounds\footnote{For the single field problem, see the discussion in Ref.~\cite{Quiros:1999jp}.}. 

Perhaps, more importantly for gauge theories at $T>0$, the standard practices for computing $\gamma_N$ are typically not gauge-invariant. At $T=0$ it is possible to adopt an expansion of the full action in powers of the gauge coupling and derivatives acting on $\phi$ that yields a gauge-invariant tunneling rate at a given order in the expansion\cite{Metaxas:1995ab}. At non-zero temperature, however, the appearance of a new scale $T$ leads to a breakdown of this power-counting approach\cite{Garny:2012cg}. In some cases, it is possible to truncate the theory at quadratic order in the gauge coupling in order to obtain a gauge-invariant $\gamma_N$ using {\em CosmoTransitions}\cite{Profumo:2014opa}. Even in this case, however, it is not clear that use of the action containing only thermal corrections to the effective potential and not the corrections to the kinetic terms in the Lagrangian is adequate. In principle, a fully non-perturbative computation of $\gamma_N(T)$ would circumvent this difficulty, though it is computationally more involved. In the interim, semi-analytic approaches that may be useful in the context of perturbation theory are being explored\cite{Espinosa:2018szu,Espinosa:2019hbm}. Further research in this direction is clearly called for. 

}

\section{Outlook}
\label{sec:out}
As the high energy physics community considers the long-term future of the energy frontier program, it is important to bear in mind any opportunities where one may -- at least in principle -- anticipate reaching definitive conclusions about the laws of nature. In this discussion, I hope to have convinced the reader that the probing the thermal history of EWSB constitutes one such opportunity. In brief: nature has handed us a scale, the electroweak temperature $\tew$. Any physics that significantly alters the SM EWSB crossover transition at this temperature cannot be arbitrarily heavy with respect to $\tew$, nor can it interact too feebly with the SM Higgs boson. \mrm{Importantly, deriving the resulting quantitative expectations for new particle masses and coupling strengths does not rely on the limit in which these particles decouple from the thermal bath. Moreover, while the SM effective field theory framework yields consistent expectations, it does not capture the full set of possible EWSB thermal histories that appear when additional degrees of freedom are included in the theory expicitly. Equally as important,} the corresponding mass scale of new particles and inferred modifications of SM Higgs boson properties appear to be within the reach of various future high energy colliders currently under consideration. An experimental program that includes one or more of these prospective colliders could conceivably teach us whether the EWSB transition was, for all intents and purposes, a crossover transition, or whether the preconditions existed for generating the matter-antimatter asymmetry in conjunction with EWSB, with possible associated astrophysical imprints in relic gravitational waves.

On the theoretical side, performing state-of-the art computations of the dynamics at non-zero temperature and making a robust correlation with the possible experimental signatures is vital. In this regard, the the work carried out starting over two decades ago on the SM EWSB transition, using the methods of high-temperature effective theory and lattice gauge theory simulations, provides a roadmap for the future (for a clear pedagogical discussions, see, {\it e.g.}, Ref.~\cite{Laine:2000xu}). In particular, the use of perturbation theory (PT) to analyze the $T>0$ behavior of gauge theory suffers from well-known limitations (see, {\it e.g.}, Refs.~\cite{Dolan:1973qd,Linde:1980ts,Gross:1980br}). While reliance on PT is unavoidable when initially assessing the EWPT implications of a wide range of explicit models, and while one may turn to various strategies to improve the performance of $T>0$ PT\mrm{\cite{Patel:2011th,Laine:1994zq,Garny:2012cg,Curtin:2016urg,Ekstedt:2018ftj,Ekstedt:2020abj}}, results of non-perturbative studies are ultimately needed to gauge the reliability of perturbative studies. Thus, a combination of perturbative analyses of BSM scenarios and non-perturbative computations intended to \lq\lq benchmark" PT appears warranted. In this respect, recent work involving high-$T$ effective theory and explicit lattice simulations\cite{Brauner:2016fla,Andersen:2017ika,Niemi:2018asa,Gould:2019qek,Kainulainen:2019kyp} is an encouraging sign.

\begin{acknowledgments}
{ I am grateful to D. Curtin, J. Gao, J. Kozaczuk, M. Mangano, P. Meade, L. Niemi, E. Senaha, J. Shelton, M. Spannowsky, S. Su, and T. V. I. Tenkanen for several useful conversations during the course of this work. I am also indebted to G. Li, J. C. Vasquez, and J. Zhou who performed several cross checks of some of the expressions appearing in this paper and for assistance with numerical evaluation of the $pp$ cross sections. This work was supported in part under U.S. Department of Energy contract DE-SC0011095 and National Science Foundation of China grant No. 19Z103010239.}
\end{acknowledgments}

\vskip 0.2in

\bibliographystyle{JHEP}
\bibliography{tewrefs_v4}

\end{document}